\documentclass[%
 aip,
 sd,%
 amsmath,amssymb,
preprint,%
]{revtex4-1}

\usepackage{multirow}
\usepackage{graphicx}
\usepackage{bm}
\usepackage{hyperref}

\usepackage{graphicx, subfigure, textcomp, amssymb, mathrsfs}

\usepackage[version=3]{mhchem} 
\usepackage{chemarr}

 \makeatletter
\newcommand\xleftrightarrow[2][]{%
  \ext@arrow 9999{\longleftrightarrowfill@}{#1}{#2}}
\newcommand\longleftrightarrowfill@{%
  \arrowfill@\leftarrow\relbar\rightarrow}
\makeatother

\usepackage{epstopdf}

\begin{document}

\title{Power-law charge relaxation of inhomogeneous porous  capacitive electrodes}

\author{Anis Allagui$^*$}
\email{aallagui@sharjah.ac.ae}
\affiliation{Dept. of Sustainable and Renewable Energy Engineering, University of Sharjah, Sharjah, P.O. Box 27272, United Arab Emirates}
\altaffiliation[Also at ]{Center for Advanced Materials Research, Research Institute of Sciences and Engineering, University of Sharjah, Sharjah,, P.O. Box 27272,  United Arab Emirates}
\affiliation{Dept. of Mechanical and Materials Engineering, Florida International University, Miami, FL33174, United States}

\author{Hachemi Benaoum} 
\affiliation{
Dept. of Applied Physics and Astronomy, 
University of Sharjah, PO Box 27272, Sharjah, United Arab Emirates 
}

\begin{abstract}

Porous electrodes--made of hierarchically nanostructured materials--are omnipresent in various electrochemical energy technologies from batteries and supercapacitors to sensors and electrocatalysis. 
Modeling   the system-level macroscopic transport and relaxation in such electrodes  given their complex microscopic geometric structure is important to better understand the performance of the devices in which they are used. 
The discharge response of capacitive porous electrodes in particular do not necessarily follow the traditional exponential decay observed with flat electrodes, which is good enough for describing the general dynamics of  processes in which the rate of a dynamic quantity (such as charge) is proportional to the quantity itself. Electric double-layer capacitors (EDLCs) and other similar systems exhibit instead power law-like discharge profiles that are best described with differential equations involving non-integer derivatives. 
Using the fractional-order integral in the Riemann-Liouville sense and superstatistics we present a treatment of the macroscopic response of such type of electrode systems starting from the mesoscopic behavior of sub-parts of it. The solutions can be in terms of the Mittag-Leffler (ML) function or a power law-like function depending on the underlying assumptions made on the physical parameters of initial charge and characteristic time response. The generalized three-parameter ML function is found to be the best suited to describe   experimental results of a commercial EDLC  at different time scales of discharge.  
 


\end{abstract}

\maketitle

\section{Introduction}

The electrodes used in many applications such as  electric double-layer and pseudocapacitive supercapacitors, capacitive deionization systems, rechargeable batteries, fuel cells, and electrochemical sensors are in general made of nanosized, multifunctional composite materials.  
Several architectures, type  one-dimensional (e.g. nanowires, nanoribbons, nanotubes), two-dimensional (e.g. nanosheets, nanoplates), and three-dimensional    materials, with doping and functionalizing additives have been designed and developed with the goal to facilitate  electrolyte transport and ion diffusion and migration, and also maximizing the amount of available atomic sites per unit  surface or volume of  the electrode \cite{liu2019three, KOSTOGLOU201749, arico2011nanostructured}. When looking at such materials in three-dimensional electrode   assembly using direct observation techniques such as optical, electron or scanning tunneling microscopies, they appear to be in the form of macroporous/mesoporous ordered or disordered interconnected channels depending of the size distribution of particles used and the preparation technique. The porosity property can cover a broad spectrum from very open porosity, such as reticulated foams, to less open structures, limited, and down to  closed porosities \cite{rice1993evaluating}. The shape of particles and their statistical  distributions functions as well as the packing density affect the   overall electrode porosity, and by deduction its performance  \cite{nimmo2004porosity}.  With these geometrical features promoting  the intimate contact of the active material with the electrolytic solution (as in batteries and supercapacitors) or a gas phase (as in fuel cells) in a compact way,  the overall device's   electrochemical performance are in fact greatly improved compared to flat electrodes \cite{newman1975porous}.  

While porous electrodes are clearly becoming ubiquitous in most electrochemical systems with considerable practical interest, they also present some challenges that need to be understood and addressed \cite{Fuller_1994, Hasyim_2017, Huang_2020, Thomas_2003}. 
On the electrode scale of modern insertion  batteries for instance, typically $10^{10}$-$10^{17}$ electrode particles of different particle sizes are present \cite{dreyer2010thermodynamic} which    lead to spatial inhomogeneities in potential and current densities \cite{orvananos2014particle, li2014current}. Localized high current densities may lead to the so-called current hotspots inducing mechanical fracture and accelerated capacity fading \cite{christensen2006stress, woodford2010electrochemical, li2014current}. In such  multiparticle electrodes the insertion does not proceed in all particles coherently (simultaneously), and as such results in distributed response times \cite{dreyer2010thermodynamic}. The same  spatial heterogeneities for the constituting materials and the resulting nonuniform local electrochemical behavior can be imagined for the case of electrodes used in electric double-layer capacitors (EDLC) or capacitive deionization systems, which are predominately made of activated carbon particles or sheets \cite{qu_charging_2018, JMCA,2017-3, ACSApplEnergyMater, 2017-3}, or for other types and configurations of electrochemical devices \cite{orgElectronics,oe2}. We note at this point that in addition to microscopy, the geometric properties of porous structures can also be indirectly assessed from the interpretation of experimental measurements of transport and relaxation processes, such as fluid flow, mercury porosimetry, electrical conduction, or small angle X-ray   and neutron scattering \cite{hilfer1996transport}. With these practical techniques, one can provide  characterization measures or parameters for porous media without specifying its porous geometry in all its possible detail \cite{Fuller_1994, Hasyim_2017, Huang_2020, C5EE00488H, C7CP00736A, prehal_salt_2018}.

The purpose of this study is to investigate the relationship between the part and the whole of a porous, blocking capacitive electrode. We are interested in the system-level electrode discharge dynamics resulting from the collective behavior of charge storage on infinitesimally  small parts of it. In other words, how the effective discharge  of a porous electrode vs. time is influenced by its microscopic geometric structure? For the reasons given above on the origins of inhomogeneities, we consider the two cases of  (i) variable independent and identically distributed (i.i.d.) initial charge accumulated on elemental surfaces of the electrode but with constant characteristic  response time for each when it is being discharged, and then (ii) the case of variable initial charge and variable characteristic response time for each elemental surface. We employ analytical techniques based on    time derivatives and integrals of non-integer order (fractional-order calculus)  \cite{haubold2000fractional, mainardi1996fractional,mathai2007pathway,mathai2009h,saxena2004generalized} and on the framework of superstatistics \cite{beck2004superstatistics, beck2003superstatistics} to transition the integral treatment of the problem from  a mesoscopic to a macroscopic size scale. While the   exponential function is the eigenfunction of the  first-order time derivative operator representing standard relaxation problems, fractional time derivatives/antiderivatives lead to solutions involving the Mittag-Leffler (ML) function.  
 We validate and  compare the derived models on data collected on a commercial EDLC exhibiting nonexponential, power-law discharging profiles. 

 \section{Theory}

\subsection{System description}

\begin{figure}[t]
\begin{center}
\includegraphics[width=2.in,angle=90]{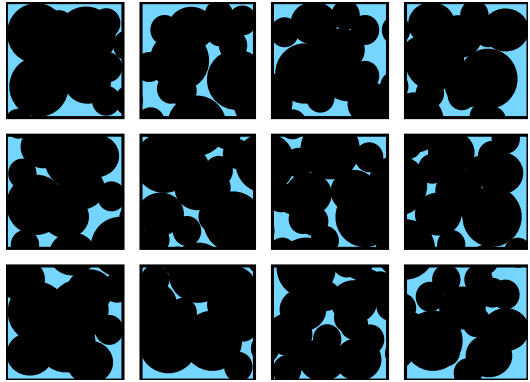}
\caption{Schematic illustration of a projection of particles of different sizes on an electrode surface,  discretized into many elemental surfaces.  The discretized surfaces contain different amounts of active material and thus different surface coverage.}
\label{fig1}
\end{center}
\end{figure}

In Fig.\;\ref{fig1} we depict schematically  a simplified geometry of a slice of a porous electrode. The porous matrices, containing a complex system of internal surfaces and phase boundaries,  are represented   by the two-dimensional projection of the active material (in black) onto the background substrate (in blue). The active material  can be composed for instance  of a single type of electronic conductor or $n$-component mixtures including essentially electronic conductors and nonconducting additives such as binders \cite{newman1975porous}. 
The electrode is discretized into a number of  elemental surfaces of constant area $\delta A_t$, as illustrated in the figure.
 Each elemental surface of the grid is covered by a different fraction of active material. 
Without going into the analysis of the exact geometric detail \cite{hilfer1996transport}, which is impractical in most cases and outside the scope of this work, we do not distinguish here between two elemental surfaces with the same amount of coverage but different geometrical arrangements of the active materials. 
 The ratio of the effective two-dimensional surface area of active material $\delta A_{a_i}$ ($i=1,2,\ldots,n$) covering an elemental surface of the electrode of area $\delta A_{\text{t}}$  is denoted by $X_i=\delta A_{a_i}/\delta A_{\text{t}} \in [0;1]$, where $A_{a_i}$ is   a random  variable. 
 We assume for simplicity that the variables $X_i$ are  i.i.d.  which means that each variable is derived from the same probability distribution function (PDF) as the other variables, and all variables are mutually independent. 
This leads to the situation in which when a potential is applied on the electrode made in contact with an electrolyte, the accumulated  charge (for a blocking electrode for example) will also be a random variable. We denote such a random variable by $Y_i=q_{0_i}/q_{\text{t}}$, where $q_{a_i}$ is the effective amount of charge  normalized with respect to $q_{\text{t}}$ which represents the case where the whole elemental surface is covered by active conducting material (i.e. maximum charge possible).    

The goal again is to derive expressions for the collective charge relaxation response of the electrode system based on the statistical heterogeneities of the microstructure.  
 
\subsection{Elemental surface response}

Consider an elemental  surface from Fig.\;\ref{fig1}  (part of the whole of an ideal capacitive electrode)  pre-charged with a constant voltage source $V_B$ to the electrical charge state $q_0 = C V_B$. Here, the charge $q_0$ can be expressed as $q_0=N_0 e$ where  $N_0$ is the number of active atomic sites with an elemental charge $e$, and the system function $C$, considered to be constant independent of time and frequency, maps voltage to charge. When the elemental capacitor is discharged into a parallel resistor of constant resistance $R$, 
 we write using Kirchoff's voltage law the first-order ordinary differential equation for the time-dependent charge $q(t)$ as:
\begin{equation}
\frac{dq(t)}{dt} =  - \lambda {q(t)}
\label{eq1}
\end{equation}
Here $\lambda=\tau^{-1}$ and $\tau=RC$ is the positive capacitive time constant of the system in units of seconds.  Eq.\;\ref{eq1} can also be written in terms of a number density $N^{*}(t)$ as 
${dN^*(t)}/{dt^*} =  -  {N^*(t)}$
where $N^{*}(t)\in [0,1]$ is equal to $N(t)$ normalized by a certain number $N_{\text{t}}$ and $t^*=\lambda t$. 
The solution to Eq.\;\ref{eq1} with the initial condition $q(t=0)=q_0$ is given by the well-known exponential decay function:
\begin{equation}
 {q}(t)=q_0 e^{-{\lambda t}}
\label{q}
\end{equation}
which is a general feature of processes in which the rate of a dynamic quantity is proportional to the quantity itself. 
 Note that if $\lambda$ is a negative constant, Eqs.\;\ref{eq1} and\;\ref{q} would describe an exponential growth process instead.

However, many electrochemical systems exhibit  dynamics  that  do not follow closely enough the exponential decay of Eq.\;\ref{q} as shown experimentally for instance by Drazer and Zanette \cite{drazer1999experimental} for the case of transport in porous samples of packed activated carbon grains, and others for supercapacitors \cite{memoryAPL}, solar cells \cite{orgElectronics}, decay of photoluminescence  \cite{dattoli2014photoluminescence} and so on. The behavior is rather  subdiffusive with power-law like tails,  
 which  is  a quite general characteristic  feature of transport in gels and porous structures,   biological media and generally disordered systems. In such systems the motion of diffusing species is often hindered by the presence of traps or obstacles or other impediments, leading to a slower-than-diffusive motions \cite{metzler2000random}. This means that at a microscopic level, in absence of an external bias, the linear  dependence of the mean squared displacement (MSD, variance of displacement) of diffusing species on   time $t$ (i.e. $ \langle \Delta x^2(t)  \rangle \sim K_1 t$, Brownian motion) is no longer  applicable, but rather  a sub-linear power-law like profile emerges, i.e. $ \langle \Delta x^2(t)  \rangle \sim K_{\alpha} t^{\alpha}$ with $0<\alpha<1$ and $K_{\alpha}$ in units of m$^2$\,s$^{-\alpha}$.   The parameter $\alpha$ here is called the temporal transport exponent and depends on the medium in which transport is taking place, and   can be either constant or variable \cite{chechkin2002retarding}. This subdiffuse behavior can be attributed in our case of electrochemical devices to many reasons including (i) the porous nature of the  electrodes  and thus some  sort of spatial restriction subjected  onto the mobile ions, and also (ii) because of internal friction forces and continuous scattering of these ions while diffusing in the supporting electrolytes \cite{ribeiro2016active}.

For the reasons outlined above on the prevalent nonexponential behavior in porous media, we extend the analysis using fractional-order calculus, which is most suited for describing anomalous transport. 
Integrating Eq.\;\ref{eq1} gives \cite{mathai2009h}:
\begin{equation}
q(t) - q_0 = - \lambda\, _0\mathrm{D}_t^{-1} q(t)
\label{eq2}
\end{equation}
where $_0\mathrm{D}_t^{-1} (= \, _0\mathrm{I}_t^{1})$ is the standard Riemann integral operator (antiderivative), 
which can be generalized to the fractional-order integral of an arbitrary order $\nu>0$ as defined by Riemann-Liouville (RL) as:
\begin{equation}
_a\mathrm{D}_t^{-\nu} f(t) = \, _a\mathrm{I}_t^{\nu} f(t)  =  \frac{1}{\Gamma(\nu)} \int\limits_a^t f(\tau) (t-\tau)^{\nu-1} d\tau, \; t>a
\label{eqRL}
\end{equation}
where $\Gamma(\cdot)$ denotes the gamma function. 
The case of zero-order integral gives $_a\mathrm{D}_t^{0} f(t) = f(t)$ (i.e. the identity operator). Eq.\;\ref{eqRL} is essentially a convolution operation of $f(\tau)$ with the algebraic kernel $K_{\nu}(t-\tau) =(t-\tau)^{\nu-1}/\Gamma (\nu)$, which can be thought of as a function relating the influence of the past on the present. In this way memory effects are introduced, which are absent in the classical case \cite{doi:10.1063/1.1806134}. Memory effects and memory trace in EDLCs have been studied in refs.  \cite{allagui2021possibility,memQ,memoryAPL}. 
  Eq.\;\ref{eq2} is  thus extended to \cite{haubold2000fractional}:
 \begin{equation}
q(t) - q_0 = - \lambda_{\nu}\, _0\mathrm{D}_t^{-\nu} q(t)
\label{eq3}
\end{equation}
where the prefactor $\lambda$ is replaced by $\lambda_{\nu}$ in units of sec$^{1/\nu}$ to maintain proper dimensionality. 
Note that for comparison with Eq.\;\ref{eq1}, applying the RL  differential operator $_0\mathrm{D}_t^{\nu}$ to Eq.\;\ref{eq3} leads   to:
\begin{equation}
_0\mathrm{D}_t^{\nu} [q(t)-q_0] = -\lambda_{\nu}\, q(t)
\end{equation} 
Fractional-order integro-differential equations are again  commonly used to describe the behavior of porous electrodes, anomalous diffusion, and in   systems exhibiting power-law dynamics \cite{memoryAPL, kosztolowicz2021subdiffusion, mainardi1996fractional, huang2005space, chechkin2002retarding, metzler2000random}. 

The solution to this problem (Eq.\;\ref{eq3}) is obtained by first applying  the Laplace transform (i.e. $\mathcal{L}\{f(t);s\}= \int_0^{\infty} f(t) e^{-st} dt$) to both sides of the equation, which  after rearranging gives:
\begin{equation}
\tilde{q}(s) = q_0 \frac{s^{-1}}{1 + \lambda_{\nu} s^{-\nu} }
\label{eq6}
\end{equation}
Using the result \cite{saxena2004generalized}:
\begin{equation}
  \int \limits_0^{\infty}  t^{\beta-1}  {E}_{\alpha,\beta}^{\gamma} \left( -at^{\alpha}\right) e^{-st} dt =  \frac{s^{-\beta}}{(1+as^{-\alpha})^{\gamma}} 
  \label{e5}
\end{equation}
where 
\begin{equation}
{E}_{\alpha,\beta}^{\gamma} ( z ) := \sum\limits_{k=0}^{\infty} \frac{(\gamma)_k}{\Gamma(\alpha k + \beta)} \frac{z^k}{k!} \quad (\alpha,\beta, \gamma \in \mathbb{C}, \mathrm{Re}({\alpha})>0)
\label{eqML}
\end{equation}
 (with $(\gamma)_k=\Gamma(\gamma+k)/\Gamma(\gamma)$, the Pochhammer symbol) is the three-parameter ML function  \cite{prabhakar1971singular}, the inverse Laplace transform applied to Eq.\;\ref{eq6} gives the time-domain charge $q(t)$ as  \cite{mathai2007pathway}:
\begin{equation}
q(t) = 
q_0   {E}_{\nu} \left(- \lambda_{\nu} t^{\nu} \right)= 
q_0 \sum\limits_{k=0}^{\infty} \frac{(-1)^k (\lambda_{\nu} t^{\nu})^k}{\Gamma(\nu k + 1)}
\label{eqML1} \end{equation}
The  charge is now a single-parameter ML function (i.e. $\alpha=\nu$, $\beta=1$ and $\gamma=1$ in Eq.\;\ref{eqML}) instead of the exponential function (Eq.\;\ref{q}), which we readily recover as $\nu\to 1$ (i.e. $E_1(-z)=e^{-z}$). The single-parameter ML function is actually an entire function, completely monotone (if and only if $0<\nu \leqslant 1$), providing a simple way of generalizing of the exponential function,  and appears naturally in problems of fractional relaxation processes and anomalous transport \cite{mainardi2014some, lemes2016generalized, de2011models}. 
Amongst its other properties,  the leading asymptotic behaviors of $q(t)$ given by Eq.\;\ref{eqML1} at the limits $t  \to 0^+$ and $t  \to +\infty$ are  respectively the power-law functions \cite{mainardi1996fractional}:
\begin{eqnarray}
\lim\limits_{\substack{t  \to 0^+}} q_0   {E}_{\nu} \left(- \lambda_{\nu} t^{\nu} \right) &=& q_0  \left[1-\frac{\lambda_{\nu} t^{\nu}}{\Gamma(1+\nu)} \right] \label{ML0} \\
\lim\limits_{\substack{t  \to +\infty}} q_0   {E}_{\nu} \left(- \lambda_{\nu} t^{\nu} \right) &=& q_0  \left[\frac{ (\lambda_{\nu} t^{\nu})^{-1}}{\Gamma(1-\nu)} \right] 
\label{MLinfty}
\end{eqnarray}
Eq.\;\ref{ML0} indicates a faster decay than the normal exponential decay at close to zero (derivative being $-q_0\nu \lambda_{\nu} t^{\nu-1}/\Gamma(1+\nu)$ which tends to $-\infty$ compared to $-q_0\lambda$ for $e^{-{\lambda t}}$), whereas the trend is inverted at infinity (Eq.\;\ref{MLinfty}) making the ML function decaying slower than to the exponential function.  

\subsection{Macroscopic electrode response}

Assume now the situation of having a number $n$ of adjacent   elemental surfaces of the capacitive electrode as depicted in Fig.\;\ref{fig1}. The time-dependent charge on each is  described by the same form given in Eq.\;\ref{eq3}, i.e.:
 \begin{align}
q_1(t) - q_{0_1} &= - \lambda_{\nu_1}\, _0\mathrm{D}_t^{-\nu} q_1(t) \nonumber \\
q_2(t) - q_{0_2} &= - \lambda_{\nu_2}\, _0\mathrm{D}_t^{-\nu} q_2(t) \nonumber \\
&\vdots \nonumber \\
q_n(t) - q_{0_n} &=- \lambda_{\nu_n}\, _0\mathrm{D}_t^{-\nu} q_n(t) 
\end{align}
We analyze first the situation where the prefactors $\lambda_{\nu_i}$  are the same for all subsystems and equal to a constant value $\lambda_{\nu}$, but different and independent initial charge $q_{0_i}$. A unique response time constant for all elemental surfaces indicate that the type of system under test, constituted of active materials in contact with the electrolytic phase,  remains unchanged and independent of spacial location on the electrode.  The justification for variable amounts of active sites holding charge is as explained above and illustrated in Fig.\;\ref{fig1}. Furthermore, the surface coverage (or amount  of active sites) and thus the amount of accumulated charge on a given elemental surface will not affect the amount of accumulated charge on another elemental surface. 
In practice, this can be imagined for  the case  of porous electrodes in supercapacitors for instance which are usually made by coating the current collectors with slurries  of polydisperse carbon particles.  Thus, these quantities can be viewed as statistically independent from each other.  
The total charge on the electrode is then the sum:
\begin{equation}
q_{\text{t}}(t) = \sum\limits_{i=1}^{n} q_i(t)
\end{equation}  
Rewriting the relations for the charges $q_i$ using 
$q_i = N_i e$, and 
knowing that the Laplace transform of the sum of independent variables is the product of the Laplace transforms \cite{mathai2017probability}, we obtain the Laplace transform of $N_{\text{t}}(t) = q_{\text{t}}(t)/e$ as:
\begin{equation}
\tilde{N}_{\text{t}}(s) =\left[\prod\limits_{i=1}^{n} \tilde{N}_{0_i}^* \right]\frac{s^{-n}}{ (1+ \lambda_{\nu}s^{-\nu})^{n} } 
\end{equation}
We then apply the inverse Laplace transform (see Eq.\;\ref{e5}) to obtain:
\begin{equation}
N_{\text{t}}(t|\lambda_{\nu}) = \bar{N}_0 t^{n-1} E_{\nu,n}^{n} (-\lambda_{\nu} t^{\nu})
\label{eq16}
\end{equation}
where $\bar{N}_0 = \prod_{i=1}^{n} N_{0_i}^*$ in units of sec$^{1/(n-1)}$. The notation $N_{\text{t}}(t|\lambda_{\nu})$ indicates that the total number density is taken at a given connstant value for $\lambda_{\nu}$. The total charge is  expressed under these assumptions in terms of a three-parameter ML function as:
\begin{equation}
q_{\text{t}}(t|\lambda_{\nu}) = \bar{q}_0 t^{n-1} E_{\nu,n}^{n} (-\lambda_{\nu} t^{\nu})
\label{ML3}
\end{equation}
  {For the limiting case of  $\nu =1$, we have    
$q_{\text{t}}(t|\lambda_{1})= \bar{q}_0 t^{n-1} e^{-{\lambda_1 t}}/\Gamma(n)$}.

We examine  now the case where in addition to variable amount of initial charge on elemental surfaces,  the prefactor $\lambda_{\nu}$  is also taken as random. Fluctuating values of $\lambda_{\nu}$   can be attributed to spacial inhomogeneities in the electrolyte and/or the constituting materials due for instance to local defects and different degrees of oxidation of the electrode material. The electrolyte, which can be viewed of as a moving boundary,  may also affect the uniformity of time responses of elemental systems.
These local differences  may lead to distributed capacitive time constants across the electrode map. As an example of  probability model for $\lambda_{\nu}$ we consider a gamma-type distribution such that: 
\begin{equation}
g(\lambda_{\nu}) = \frac{z^n}{\Gamma(n)} \lambda_{\nu}^{n-1} e^{-  \lambda_{\nu} z}\quad ( \lambda_{\nu}>0, z>0, n>0)
\label{eqg}
\end{equation} 
which has a mean value of  $n/z$ for.  
While it is difficult to get a good estimate for the local variabilities of  $\lambda_{\nu}$,  the gamma distribution  is taken here as a possible spreading function because it is defined for positive  variables only which is close to physical systems  ($\lambda_{\nu} > 0$). 
 Furthermore, for  $n=1$, one retrieves the exponential distribution, but a number of other distributions can be obtained as special cases, such as the chi-square, Weibull,  hydrograph,  Rayleigh  or the Maxwell molecular velocity distributions \cite{lienhard1967physical}. This makes the gamma distribution versatile enough to describe many different types of statistics  \cite{beck2006stretched, beck2004superstatistics, beck2003superstatistics, mathai2012pathway, allagui2021gouy}.  
 The integral of the conditional probability expression given by Eq.\;\ref{eq16}, knowing the distribution  $g(\lambda_{\nu})$ given by Eq.\;\ref{eqg}, provides the unconditional number $N_{\text{t}}(t)$ in closed form as  \cite{saxena2004generalized,mathai2007pathway}:
 \begin{align}
N_{\text{t}}(t) &= \int\limits_{0}^{\infty}  N_{\text{t}}(t|\lambda_{\nu})   g(\lambda_{\nu})  d \lambda_{\nu}\\
&= \frac{\bar{N}_0}{\Gamma(n)} t^{n-1} \left[ 1 + \left(\frac{t}{z}\right)^{\nu} \right]^{-n}
\end{align}
 The  total charge accumulated on the electrode is then given by:
 \begin{equation}
q_{\text{t}}(t) 
= \frac{\bar{q}_0}{\Gamma(n)} t^{n-1} \left[ 1 + \left(\frac{t}{z}\right)^{\nu} \right]^{-n}
\label{PL}
\end{equation}
The limiting cases for which $t\to 0^+$ and $t\to\infty$ provide the asymptotic behaviors $t^{n-1}$ and $t^{n (\nu +1) -1}$, respectively. 
 This framework in which the statistics of the statistics is estimated is derived from the work of Beck \cite{beck2004superstatistics} and Beck and Cohen  \cite{beck2003superstatistics} on superstatistics.  It is also commonly known in studies of anomalous relaxation and transport as the subordination formalism \cite{stanislavsky2007stochastic, de2011models, stanislavsky2010subordination, chechkin2017brownian}.

\section{Experimental}

\begin{figure}[t]
\begin{center}
\subfigure[]{\includegraphics[width=1.65in]{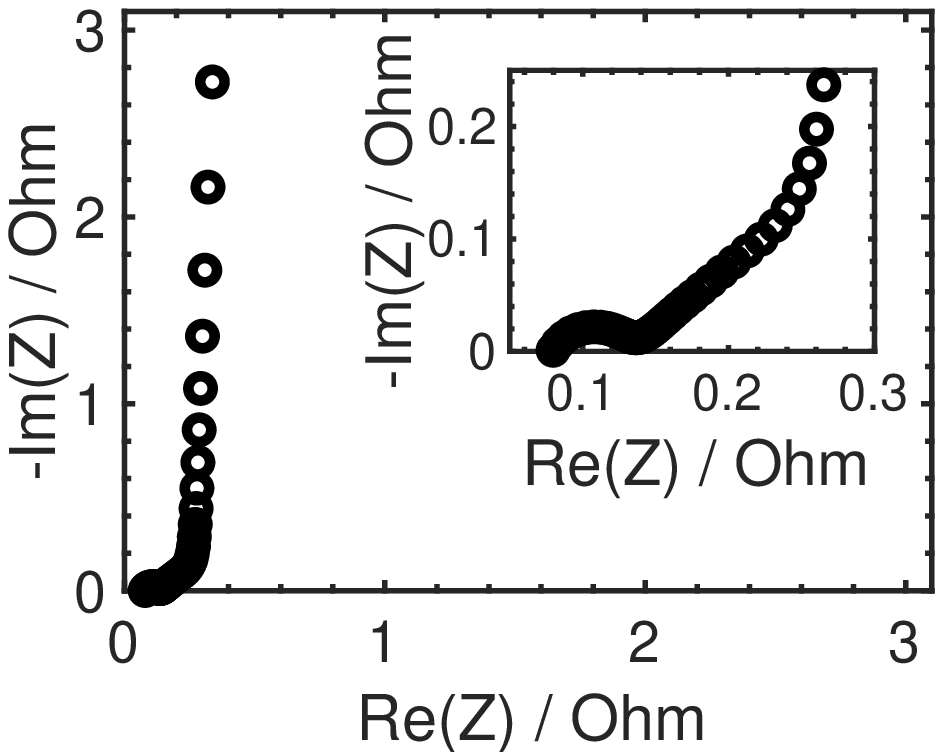}} 
\subfigure[]{\includegraphics[width=1.65in]{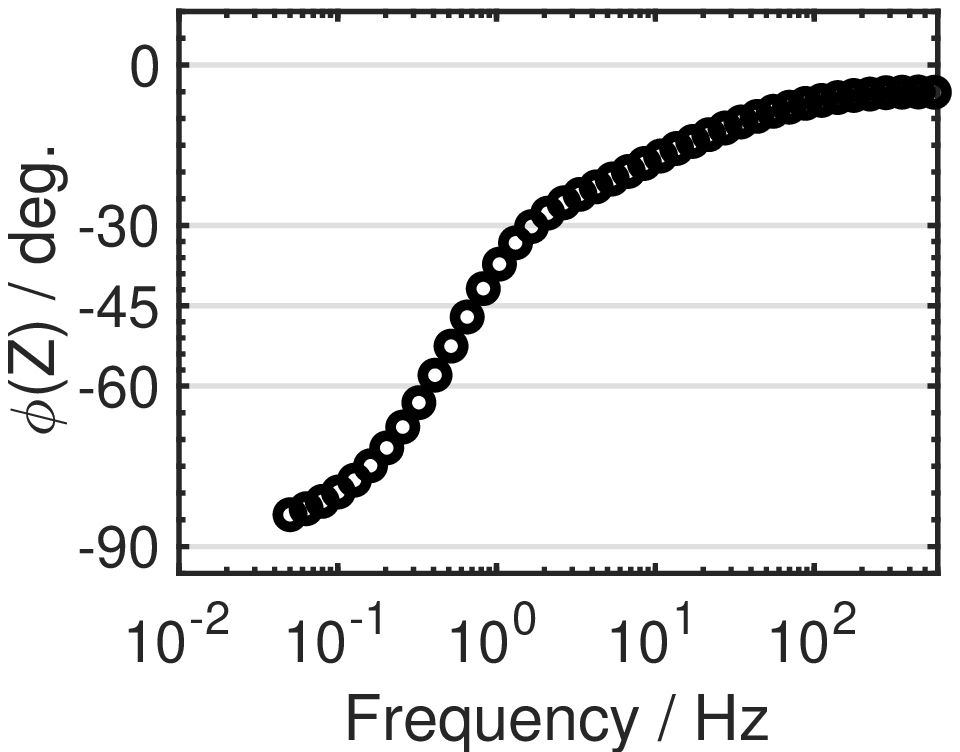}}
\caption{Impedance spectroscopy results of the Samxon EDLC (DRL series, part No. DRL105S0TF12RR, rated 2.7\,V, 1\,F) measured using sinusoidal low voltage signals at open-circuit voltage (similar results were obtained at 1.35\;Vdc and 2.70\;Vdc, but not shown here): in (a) we show the Nyquist plot of imaginary vs. real parts of impedance, and in (b) the impedance phase angle vs. frequency}
\label{figEIS}
\end{center}
\end{figure}
 
To evaluate the models given by 
Eq.\;\ref{q} (exponential function), Eq.\;\ref{eqML1} (single-parameter ML function), Eq.\;\ref{ML3} (three-parameter ML function), and Eq.\;\ref{PL} (power-law function), 
electrical measurements were carried out on a commercial electric double-layer capacitor (EDLC)    (Samxon, part No. DRL105S0TF12RR, rated 2.7\,V, 1\,F) using a Biologic VSP 300 potentiostat equipped with impedance spectroscopy module. The device's spectral impedance  results  are shown in Fig.\;\ref{figEIS}. The non-vertical  plot of imaginary vs. real parts  (Fig.\;\ref{figEIS}(a)) or the deviation of   phase angle  from -90 deg. (Fig.\;\ref{figEIS}(b))    indicate the non-ideal capacitive-resistive performance of the EDLC under test.  Detailed analysis of spectral impedance of EDLCs can be found in refs. \cite{fracorderreview,eis,memoryAPL}. 
Charge and discharge measurements were conducted as follows. First, the device was pre-charged with constant current-constant voltage (CCCV) mode: 20\,mA up to the nominal voltage of  2.7\,V, and then the voltage was maintained   at 2.7\,V for 5 minutes. For the subsequent  discharge step, the potentiostat acted as a constant resistor $R$ for the duration  necessary for the voltage to drop from 2.7\,V to 3\,mV. Four cycles of  charge and discharge were   conducted with different values of $R$ (50\,$\Omega$, 10\,$\Omega$, 2\,$\Omega$ and 0.7\,$\Omega$). The  time step for voltage and current measurements  was fixed to 50\,ms.
  
\section{Results and discussion}
 
\begin{figure}[t]
\begin{center}
\subfigure[]{\includegraphics[width=1.65in]{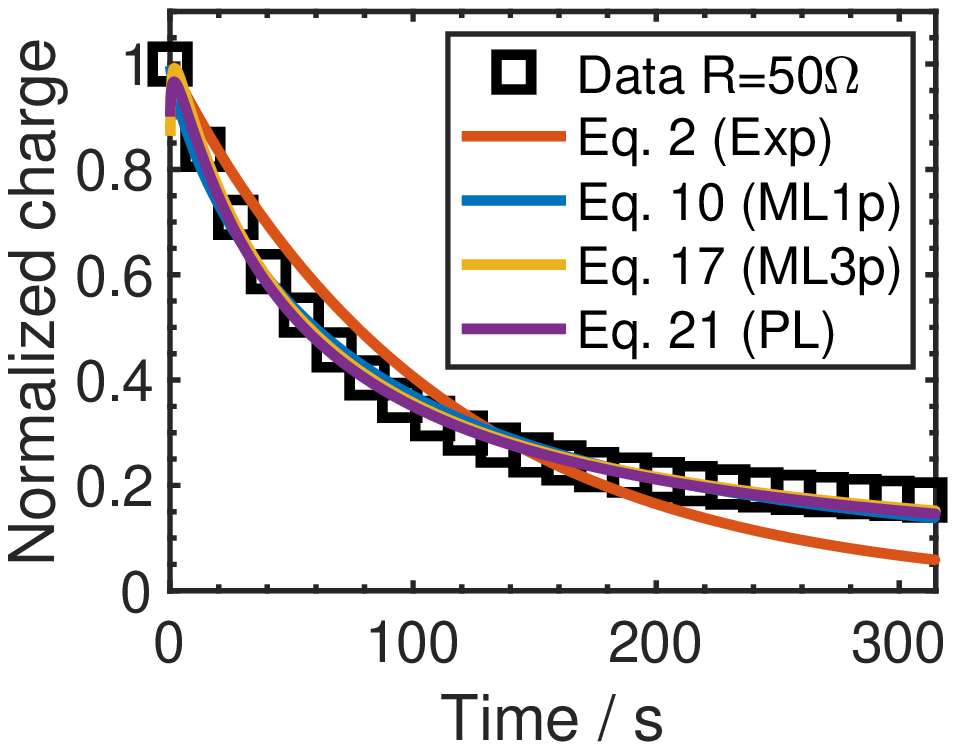}} 
\subfigure[]{\includegraphics[width=1.65in]{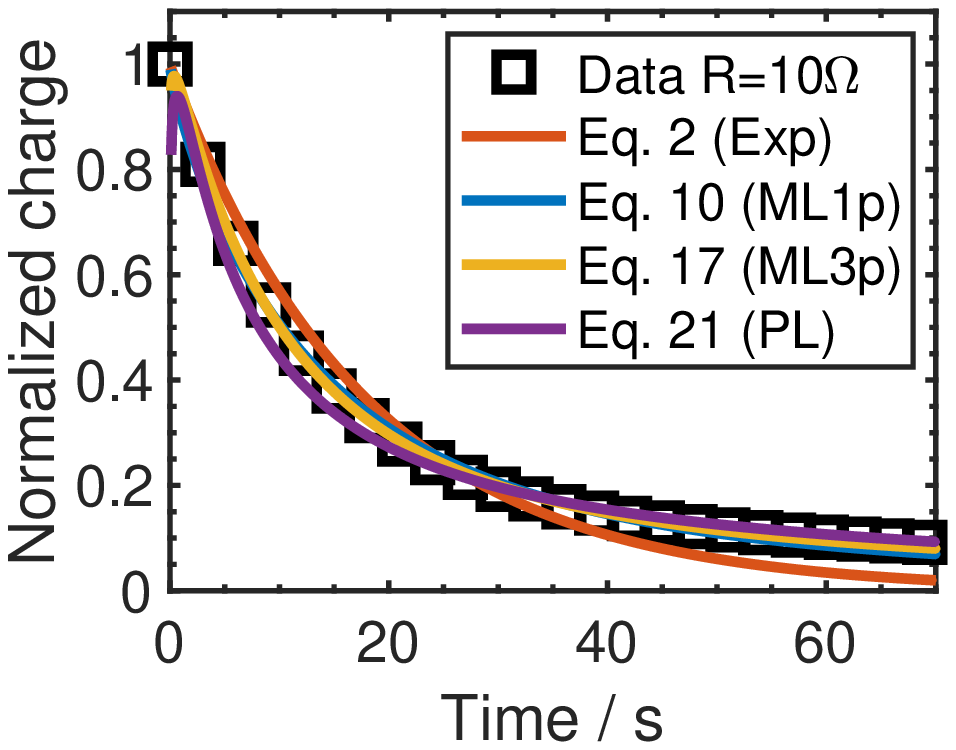}}\\
\subfigure[]{\includegraphics[width=1.65in]{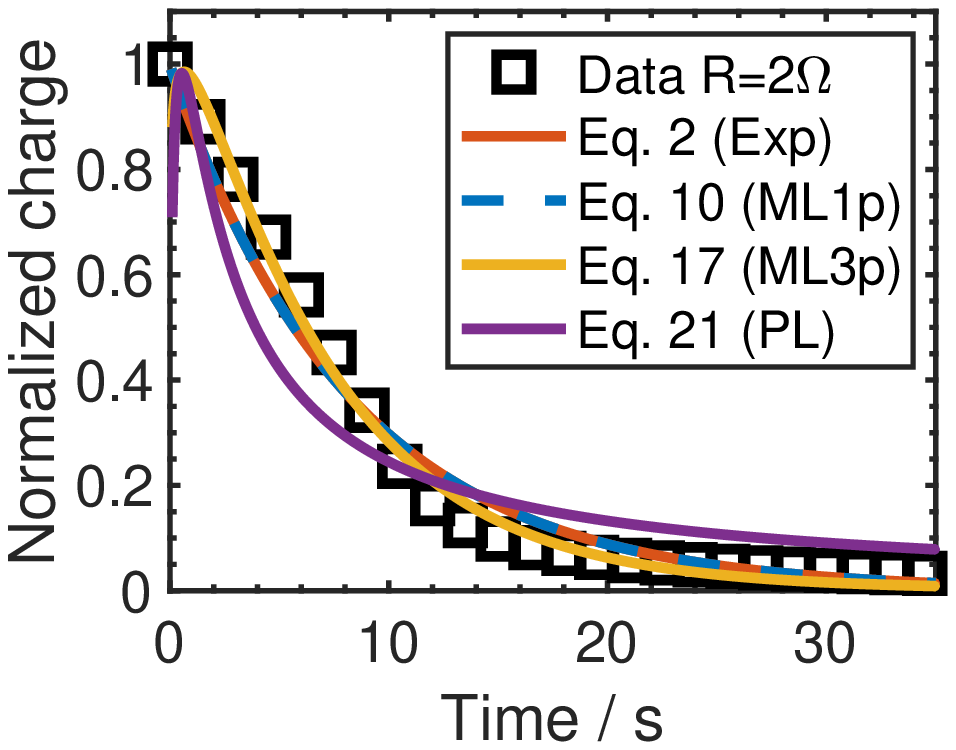}}
\subfigure[]{\includegraphics[width=1.65in]{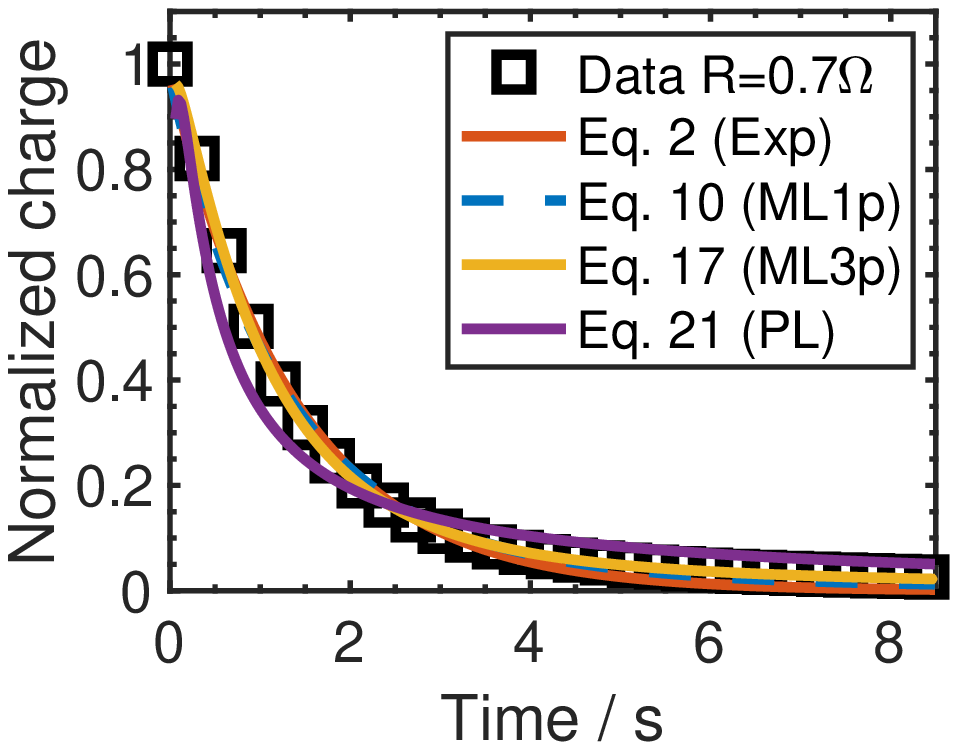}}
\caption{Plots of normalized electric charge vs. time of the Samxon EDLC (DRL series, part No. DRL105S0TF12RR, rated 2.7\,V, 1\,F) for the four cases of constant load discharge, (a) with $R=50\,\Omega$, (b) $R=10\,\Omega$, (c) $R=2\,\Omega$ and (d) $R=0.7\,\Omega$.  Nonlinear least squares data fitting  with the  model  functions    given by Eq.\;\ref{q} (denoted Exp for exponential decay), Eq.\;\ref{eqML1} (denoted ML1p for single-parameter ML function), Eq.\;\ref{ML3} (denoted ML3p for three-parameter ML function), and Eq.\;\ref{PL} (denoted PL for power-law decay) are also shown. The fitting parameters for each case and each model are summarized in Table.\;\ref{tab1}.}
\label{fig2}
\end{center}
\end{figure}

 In Fig.\;\ref{fig2}(a)  we show the measured electrical charge (normalized with its maximum) vs. time of the EDLC when discharged into a 50\,$\Omega$ resistor. Not all data points are plotted for better clarity of the figure. Results of nonlinear least squares fitting minimizing the sum of squared differences $\sum_i (y_{\text{mdl}}(t) - y_{\text{exp}}(t))^2$ where $y_{\text{exp}}(t)$ is the data and $y_{\text{mdl}}(t)$ is the model  function    given by Eqs.\;\ref{q}, \ref{eqML1}, \ref{ML3}, \ref{PL} are also shown. The same is repeated for three other values of the resistance $R$, i.e. 10\,$\Omega$, 2\,$\Omega$ and 0.7\,$\Omega$, and the results are   provided in Figs.\;\ref{fig2}(b)-(d), respectively. 
 The sets of fitting parameters for each discharge experiment and for each model are summarized in Table.\;\ref{tab1}. The   squared 2-norms of the residuals are also reported that we used as   an  optimality criterion in comparing between the different models. 
 
 It is clear from Figs.\;\ref{fig2}(a)-(b) and the corresponding values of squared norm of the residuals in Table.\;\ref{tab1} that the standard exponential function does not properly fit the experimental decaying charge of the device when compared to the performance of the two ML functions-based models (Eqs.\;\ref{eqML1} and\;\ref{ML3}) or the power-law model (Eq.\;\ref{PL}). The data for these cases show first a faster decay than the exponential function, and then a change of trend towards a slower asymptotic at longer times. Such a behavior is nicely captured  by the nonexponential models, with the values of fractional exponent  different from one.  
 The asymptotic approximations given in Eqs.\ref{MLinfty} and\;\ref{ML0} for the decaying ML function matches with  (i)     a stretched exponential as $t\to 0^+$ and thus the very fast decay,   and (ii)   a  negative power law as $t\to \infty$ and thus the very slow decay.  (iii)  For intermediate times, the ML function interpolates between the two behaviors \cite{mainardi2014some}. 
 As indicated in Table\;\ref{tab1}, the residual sum of squares are 2.690, 0.836, and 0.863 for these three non-exponential models (Eqs.\;\ref{eqML1}, \;\ref{ML3}), and \;\ref{PL}, respectively) compared to 30.71 for the exponential decay when the discharging resistor $R$ was set to 50\,$\Omega$, and     0.302, 0.067, 0.599 vs. 3.313 for $R=10$\,$\Omega$.

 \begin{table}[t]
\caption{Fitting parameters  for each discharge experiment ($R=50\,\Omega, 10\,\Omega, 2\,\Omega, 0.7\,\Omega$) using the four models of Eqs.\;\ref{q},\;\ref{eqML1},\;\ref{ML3}, and\;\ref{PL} with their squared 2-norm of the residuals. The corresponding plots of charge vs. time are provided in Fig.\;\ref{fig2}}
\begin{center}
\begin{tabular}{l | l| l | l | l}
\hline 
&   Eq.\;\ref{q}   & Eq.\;\ref{eqML1} & Eq.\;\ref{ML3}  & Eq.\;\ref{PL} \\ \hline
$R=50\,\Omega$   & $\lambda=0.009$ & $\lambda_{\nu}=0.028$  & $\lambda_{\nu}=0.066$ & $z=44.18$  \\
&  &  $\nu=0.788$ & $\nu=0.713$  & $\nu=0.977$  \\ 
&  & & $n=1.081$ &  $n=1.039$ \\ 
&  & & $\bar{q}_0=1.012$ & $\bar{q}_0=0.987$ \\
 Res. norm.&  30.71  & 2.690 &  0.836 &  0.863\\ \hline
$R=10\,\Omega$ &$\lambda=0.056   $ &$\lambda_{\nu}=0.094$& $\lambda_{\nu}=   0.147$& $z=   5.007$\\
&         &    $\nu=0.863$&  $\nu=  0.814$ &$\nu=   1.000$ \\
&         &         &   $n= 1.057$ & $n=   1.129$\\
&          &         &  $\bar{q}_0= 1.076$ & $\bar{q}_0=  1.139$ \\
 Res. norm.&   3.313&    0.302&    0.067&    0.599
\\ \hline
 $R=2\,\Omega$  & $\lambda=0.121$ &  $\lambda_{\nu}= 0.121 $ & $\lambda_{\nu}= 0.170 $ &  $z=1.208$ \\
    &       & $\nu= 1.000$   & $\nu=0.991$   & $\nu=1.000$ \\
     &      &         &$n=1.111$    &$n=1.441$ \\
      &       &        &$\bar{q}_0=1.094$    &$\bar{q}_0=2.196$ \\
      Res. nor.&     1.551&    1.551&    1.558&    5.564
\\ \hline
  $R=0.7\,\Omega$ &   $\lambda= 0.724$ &  $\lambda_{\nu}= 0.759$ &  $\lambda_{\nu}= 1.052$ & $z=  0.168$\\
   &       &   $\nu=0.961$ &  $\nu= 0.915$  & $\nu= 1.000$ \\
    &      &        &  $n= 1.075$  &  $n=1.577$ \\
     &     &         & $\bar{q}_0=  1.242 $ & $\bar{q}_0= 7.365$\\
     Res. norm. &     0.122   & 0.073  &  0.005 &   0.574
 \\ \hline

\end{tabular}
\end{center}
\label{tab1}
\end{table}%

From Figs.\;\ref{fig2}(c)-(d), however, in which the parallel resistances are smaller and thus the discharging is faster, the data do not show enough of the fat tail at longer times.  The goodness-of-fit of the exponential decay and the two  ML functions-based models are comparable for the results obtained with $R=2\,\Omega$, but the   best fitting one for the data of $R=0.7\,\Omega$ is the three-parameter ML model with the lowest values of residual sum of squares (see Table.\;\ref{tab1}). The power law model given by Eq.\;\ref{PL} seems to be the one that deviates the most from the experiments. Recall that it is derived from the collective response of  subsystems  of variable time constants and variable initial charge. Despite the fact that this model has four free parameters like the case of the generalized three-parameter ML model, its fitting performance is still inferior. This can be attributed to the effect of the chosen  PDF to describe the spreading of $\lambda_{nu}$.  
Other PDFs may provide better fits, knowing however that it is experimentally difficult to estimate locally such variabilities. 

\section{Conclusion} 
In this study we proposed a way of describing the macroscopic charge relaxation behavior of an EDLC constituted of porous electrodes from a mesoscopic description of the charge dynamics on   elemental parts of it. Because of the inherent inhomogeneities associated with porous media, we first assumed the initial charges accumulated on elemental subsystems to be derived from a random   distribution which led to a solution in terms of the three-parameter ML function. Such a model successfully captured with great fidelity  the discharge behavior  of the EDLC device for different values of the applied resistive load. When integrating furthermore a variable time constant of these subsystems (derived from a gamma PDF), the resulting power-law model did not perform well enough for fast discharging situations. We concluded that the three-parameter ML model, which is able to span the behavior of the stretched exponential function at small values of time as well as  that of a negative power law  as time grows, is more suited for describing the dynamics of porous electrodes at different time scales. 
 In principle the same analysis can be adapted to similar  systems (porous and heterogeneous) involving the transport or relaxation of some other physical quantities such as energy, momentum or mass. 

\section*{Acknowledgement}

A.A. acknowledges the support provided by the NSF, project \#2126190.



\begin{thebibliography}{57}%
\makeatletter
\providecommand \@ifxundefined [1]{%
 \@ifx{#1\undefined}
}%
\providecommand \@ifnum [1]{%
 \ifnum #1\expandafter \@firstoftwo
 \else \expandafter \@secondoftwo
 \fi
}%
\providecommand \@ifx [1]{%
 \ifx #1\expandafter \@firstoftwo
 \else \expandafter \@secondoftwo
 \fi
}%
\providecommand \natexlab [1]{#1}%
\providecommand \enquote  [1]{``#1''}%
\providecommand \bibnamefont  [1]{#1}%
\providecommand \bibfnamefont [1]{#1}%
\providecommand \citenamefont [1]{#1}%
\providecommand \href@noop [0]{\@secondoftwo}%
\providecommand \href [0]{\begingroup \@sanitize@url \@href}%
\providecommand \@href[1]{\@@startlink{#1}\@@href}%
\providecommand \@@href[1]{\endgroup#1\@@endlink}%
\providecommand \@sanitize@url [0]{\catcode `\\12\catcode `\$12\catcode
  `\&12\catcode `\#12\catcode `\^12\catcode `\_12\catcode `\%12\relax}%
\providecommand \@@startlink[1]{}%
\providecommand \@@endlink[0]{}%
\providecommand \url  [0]{\begingroup\@sanitize@url \@url }%
\providecommand \@url [1]{\endgroup\@href {#1}{\urlprefix }}%
\providecommand \urlprefix  [0]{URL }%
\providecommand \Eprint [0]{\href }%
\providecommand \doibase [0]{http://dx.doi.org/}%
\providecommand \selectlanguage [0]{\@gobble}%
\providecommand \bibinfo  [0]{\@secondoftwo}%
\providecommand \bibfield  [0]{\@secondoftwo}%
\providecommand \translation [1]{[#1]}%
\providecommand \BibitemOpen [0]{}%
\providecommand \bibitemStop [0]{}%
\providecommand \bibitemNoStop [0]{.\EOS\space}%
\providecommand \EOS [0]{\spacefactor3000\relax}%
\providecommand \BibitemShut  [1]{\csname bibitem#1\endcsname}%
\let\auto@bib@innerbib\@empty
\bibitem [{\citenamefont {Liu}\ \emph {et~al.}(2019)\citenamefont {Liu},
  \citenamefont {Yuan}, \citenamefont {Zhang}, \citenamefont {Wang},
  \citenamefont {Huang}, \citenamefont {Yu}, \citenamefont {Zhu}, \citenamefont
  {Fu}, \citenamefont {Wang}, \citenamefont {Chen},\ and\ \citenamefont
  {Wu}}]{liu2019three}%
  \BibitemOpen
  \bibfield  {author} {\bibinfo {author} {\bibfnamefont {Z.}~\bibnamefont
  {Liu}}, \bibinfo {author} {\bibfnamefont {X.}~\bibnamefont {Yuan}}, \bibinfo
  {author} {\bibfnamefont {S.}~\bibnamefont {Zhang}}, \bibinfo {author}
  {\bibfnamefont {J.}~\bibnamefont {Wang}}, \bibinfo {author} {\bibfnamefont
  {Q.}~\bibnamefont {Huang}}, \bibinfo {author} {\bibfnamefont
  {N.}~\bibnamefont {Yu}}, \bibinfo {author} {\bibfnamefont {Y.}~\bibnamefont
  {Zhu}}, \bibinfo {author} {\bibfnamefont {L.}~\bibnamefont {Fu}}, \bibinfo
  {author} {\bibfnamefont {F.}~\bibnamefont {Wang}}, \bibinfo {author}
  {\bibfnamefont {Y.}~\bibnamefont {Chen}}, \ and\ \bibinfo {author}
  {\bibfnamefont {Y.}~\bibnamefont {Wu}},\ }\bibfield  {title} {\enquote
  {\bibinfo {title} {Three-dimensional ordered porous electrode materials for
  electrochemical energy storage},}\ }\href@noop {} {\bibfield  {journal}
  {\bibinfo  {journal} {NPG Asia Materials}\ }\textbf {\bibinfo {volume}
  {11}},\ \bibinfo {pages} {1--21} (\bibinfo {year} {2019})}\BibitemShut
  {NoStop}%
\bibitem [{\citenamefont {Kostoglou}\ \emph {et~al.}(2017)\citenamefont
  {Kostoglou}, \citenamefont {Koczwara}, \citenamefont {Prehal}, \citenamefont
  {Terziyska}, \citenamefont {Babic}, \citenamefont {Matovic}, \citenamefont
  {Constantinides}, \citenamefont {Tampaxis}, \citenamefont {Charalambopoulou},
  \citenamefont {Steriotis}, \citenamefont {Hinder}, \citenamefont {Baker},
  \citenamefont {Polychronopoulou}, \citenamefont {Doumanidis}, \citenamefont
  {Paris}, \citenamefont {Mitterer},\ and\ \citenamefont
  {Rebholz}}]{KOSTOGLOU201749}%
  \BibitemOpen
  \bibfield  {author} {\bibinfo {author} {\bibfnamefont {N.}~\bibnamefont
  {Kostoglou}}, \bibinfo {author} {\bibfnamefont {C.}~\bibnamefont {Koczwara}},
  \bibinfo {author} {\bibfnamefont {C.}~\bibnamefont {Prehal}}, \bibinfo
  {author} {\bibfnamefont {V.}~\bibnamefont {Terziyska}}, \bibinfo {author}
  {\bibfnamefont {B.}~\bibnamefont {Babic}}, \bibinfo {author} {\bibfnamefont
  {B.}~\bibnamefont {Matovic}}, \bibinfo {author} {\bibfnamefont
  {G.}~\bibnamefont {Constantinides}}, \bibinfo {author} {\bibfnamefont
  {C.}~\bibnamefont {Tampaxis}}, \bibinfo {author} {\bibfnamefont
  {G.}~\bibnamefont {Charalambopoulou}}, \bibinfo {author} {\bibfnamefont
  {T.}~\bibnamefont {Steriotis}}, \bibinfo {author} {\bibfnamefont
  {S.}~\bibnamefont {Hinder}}, \bibinfo {author} {\bibfnamefont
  {M.}~\bibnamefont {Baker}}, \bibinfo {author} {\bibfnamefont
  {K.}~\bibnamefont {Polychronopoulou}}, \bibinfo {author} {\bibfnamefont
  {C.}~\bibnamefont {Doumanidis}}, \bibinfo {author} {\bibfnamefont
  {O.}~\bibnamefont {Paris}}, \bibinfo {author} {\bibfnamefont
  {C.}~\bibnamefont {Mitterer}}, \ and\ \bibinfo {author} {\bibfnamefont
  {C.}~\bibnamefont {Rebholz}},\ }\bibfield  {title} {\enquote {\bibinfo
  {title} {Nanoporous activated carbon cloth as a versatile material for
  hydrogen adsorption, selective gas separation and electrochemical energy
  storage},}\ }\href {\doibase https://doi.org/10.1016/j.nanoen.2017.07.056}
  {\bibfield  {journal} {\bibinfo  {journal} {Nano Energy}\ }\textbf {\bibinfo
  {volume} {40}},\ \bibinfo {pages} {49--64} (\bibinfo {year}
  {2017})}\BibitemShut {NoStop}%
\bibitem [{\citenamefont {Arico}\ \emph {et~al.}(2011)\citenamefont {Arico},
  \citenamefont {Bruce}, \citenamefont {Scrosati}, \citenamefont {Tarascon},\
  and\ \citenamefont {Van~Schalkwijk}}]{arico2011nanostructured}%
  \BibitemOpen
  \bibfield  {author} {\bibinfo {author} {\bibfnamefont {A.~S.}\ \bibnamefont
  {Arico}}, \bibinfo {author} {\bibfnamefont {P.}~\bibnamefont {Bruce}},
  \bibinfo {author} {\bibfnamefont {B.}~\bibnamefont {Scrosati}}, \bibinfo
  {author} {\bibfnamefont {J.-M.}\ \bibnamefont {Tarascon}}, \ and\ \bibinfo
  {author} {\bibfnamefont {W.}~\bibnamefont {Van~Schalkwijk}},\ }\bibfield
  {title} {\enquote {\bibinfo {title} {Nanostructured materials for advanced
  energy conversion and storage devices},}\ }\href@noop {} {\bibfield
  {journal} {\bibinfo  {journal} {Materials for sustainable energy: a
  collection of peer-reviewed research and review articles from Nature
  Publishing Group}\ ,\ \bibinfo {pages} {148--159}} (\bibinfo {year}
  {2011})}\BibitemShut {NoStop}%
\bibitem [{\citenamefont {Rice}(1993)}]{rice1993evaluating}%
  \BibitemOpen
  \bibfield  {author} {\bibinfo {author} {\bibfnamefont {R.~W.}\ \bibnamefont
  {Rice}},\ }\bibfield  {title} {\enquote {\bibinfo {title} {Evaluating
  porosity parameters for porosity--property relations},}\ }\href@noop {}
  {\bibfield  {journal} {\bibinfo  {journal} {Journal of the American Ceramic
  Society}\ }\textbf {\bibinfo {volume} {76}},\ \bibinfo {pages} {1801--1808}
  (\bibinfo {year} {1993})}\BibitemShut {NoStop}%
\bibitem [{\citenamefont {Nimmo}(2004)}]{nimmo2004porosity}%
  \BibitemOpen
  \bibfield  {author} {\bibinfo {author} {\bibfnamefont {J.~R.}\ \bibnamefont
  {Nimmo}},\ }\bibfield  {title} {\enquote {\bibinfo {title} {Porosity and pore
  size distribution},}\ }\href@noop {} {\bibfield  {journal} {\bibinfo
  {journal} {Encyclopedia of Soils in the Environment}\ }\textbf {\bibinfo
  {volume} {3}},\ \bibinfo {pages} {295--303} (\bibinfo {year}
  {2004})}\BibitemShut {NoStop}%
\bibitem [{\citenamefont {Newman}\ and\ \citenamefont
  {Tiedemann}(1975)}]{newman1975porous}%
  \BibitemOpen
  \bibfield  {author} {\bibinfo {author} {\bibfnamefont {J.}~\bibnamefont
  {Newman}}\ and\ \bibinfo {author} {\bibfnamefont {W.}~\bibnamefont
  {Tiedemann}},\ }\bibfield  {title} {\enquote {\bibinfo {title}
  {Porous-electrode theory with battery applications},}\ }\href@noop {}
  {\bibfield  {journal} {\bibinfo  {journal} {AIChE Journal}\ }\textbf
  {\bibinfo {volume} {21}},\ \bibinfo {pages} {25--41} (\bibinfo {year}
  {1975})}\BibitemShut {NoStop}%
\bibitem [{\citenamefont {Fuller}, \citenamefont {Doyle},\ and\ \citenamefont
  {Newman}(1994)}]{Fuller_1994}%
  \BibitemOpen
  \bibfield  {author} {\bibinfo {author} {\bibfnamefont {T.~F.}\ \bibnamefont
  {Fuller}}, \bibinfo {author} {\bibfnamefont {M.}~\bibnamefont {Doyle}}, \
  and\ \bibinfo {author} {\bibfnamefont {J.}~\bibnamefont {Newman}},\
  }\bibfield  {title} {\enquote {\bibinfo {title} {Relaxation phenomena in
  lithium-ion-insertion cells},}\ }\href {\doibase 10.1149/1.2054868}
  {\bibfield  {journal} {\bibinfo  {journal} {J. Electrochem. Soc.}\ }\textbf
  {\bibinfo {volume} {141}},\ \bibinfo {pages} {982--990} (\bibinfo {year}
  {1994})}\BibitemShut {NoStop}%
\bibitem [{\citenamefont {Hasyim}\ \emph {et~al.}(2017)\citenamefont {Hasyim},
  \citenamefont {Ma}, \citenamefont {Rajagopalan},\ and\ \citenamefont
  {Randall}}]{Hasyim_2017}%
  \BibitemOpen
  \bibfield  {author} {\bibinfo {author} {\bibfnamefont {M.~R.}\ \bibnamefont
  {Hasyim}}, \bibinfo {author} {\bibfnamefont {D.}~\bibnamefont {Ma}}, \bibinfo
  {author} {\bibfnamefont {R.}~\bibnamefont {Rajagopalan}}, \ and\ \bibinfo
  {author} {\bibfnamefont {C.}~\bibnamefont {Randall}},\ }\bibfield  {title}
  {\enquote {\bibinfo {title} {Prediction of charge-discharge and impedance
  characteristics of electric double-layer capacitors using porous electrode
  theory},}\ }\href {\doibase 10.1149/2.0051713jes} {\bibfield  {journal}
  {\bibinfo  {journal} {J. Electrochem. Soc.}\ }\textbf {\bibinfo {volume}
  {164}},\ \bibinfo {pages} {A2899--A2913} (\bibinfo {year}
  {2017})}\BibitemShut {NoStop}%
\bibitem [{\citenamefont {Huang}\ \emph {et~al.}(2020)\citenamefont {Huang},
  \citenamefont {Gao}, \citenamefont {Luo}, \citenamefont {Wang}, \citenamefont
  {Li}, \citenamefont {Chen},\ and\ \citenamefont {Zhang}}]{Huang_2020}%
  \BibitemOpen
  \bibfield  {author} {\bibinfo {author} {\bibfnamefont {J.}~\bibnamefont
  {Huang}}, \bibinfo {author} {\bibfnamefont {Y.}~\bibnamefont {Gao}}, \bibinfo
  {author} {\bibfnamefont {J.}~\bibnamefont {Luo}}, \bibinfo {author}
  {\bibfnamefont {S.}~\bibnamefont {Wang}}, \bibinfo {author} {\bibfnamefont
  {C.}~\bibnamefont {Li}}, \bibinfo {author} {\bibfnamefont {S.}~\bibnamefont
  {Chen}}, \ and\ \bibinfo {author} {\bibfnamefont {J.}~\bibnamefont {Zhang}},\
  }\bibfield  {title} {\enquote {\bibinfo {title} {Impedance response of porous
  electrodes: Theoretical framework, physical models and applications},}\
  }\href {\doibase 10.1149/1945-7111/abc655} {\bibfield  {journal} {\bibinfo
  {journal} {J. Electrochem. Soc.}\ }\textbf {\bibinfo {volume} {167}},\
  \bibinfo {pages} {166503} (\bibinfo {year} {2020})}\BibitemShut {NoStop}%
\bibitem [{\citenamefont {Thomas}\ and\ \citenamefont
  {Newman}(2003)}]{Thomas_2003}%
  \BibitemOpen
  \bibfield  {author} {\bibinfo {author} {\bibfnamefont {K.~E.}\ \bibnamefont
  {Thomas}}\ and\ \bibinfo {author} {\bibfnamefont {J.}~\bibnamefont
  {Newman}},\ }\bibfield  {title} {\enquote {\bibinfo {title} {Thermal modeling
  of porous insertion electrodes},}\ }\href {\doibase 10.1149/1.1531194}
  {\bibfield  {journal} {\bibinfo  {journal} {J. Electrochem. Soc.}\ }\textbf
  {\bibinfo {volume} {150}},\ \bibinfo {pages} {A176} (\bibinfo {year}
  {2003})}\BibitemShut {NoStop}%
\bibitem [{\citenamefont {Dreyer}\ \emph {et~al.}(2010)\citenamefont {Dreyer},
  \citenamefont {Jamnik}, \citenamefont {Guhlke}, \citenamefont {Huth},
  \citenamefont {Mo{\v{s}}kon},\ and\ \citenamefont
  {Gaber{\v{s}}{\v{c}}ek}}]{dreyer2010thermodynamic}%
  \BibitemOpen
  \bibfield  {author} {\bibinfo {author} {\bibfnamefont {W.}~\bibnamefont
  {Dreyer}}, \bibinfo {author} {\bibfnamefont {J.}~\bibnamefont {Jamnik}},
  \bibinfo {author} {\bibfnamefont {C.}~\bibnamefont {Guhlke}}, \bibinfo
  {author} {\bibfnamefont {R.}~\bibnamefont {Huth}}, \bibinfo {author}
  {\bibfnamefont {J.}~\bibnamefont {Mo{\v{s}}kon}}, \ and\ \bibinfo {author}
  {\bibfnamefont {M.}~\bibnamefont {Gaber{\v{s}}{\v{c}}ek}},\ }\bibfield
  {title} {\enquote {\bibinfo {title} {The thermodynamic origin of hysteresis
  in insertion batteries},}\ }\href@noop {} {\bibfield  {journal} {\bibinfo
  {journal} {Nature materials}\ }\textbf {\bibinfo {volume} {9}},\ \bibinfo
  {pages} {448--453} (\bibinfo {year} {2010})}\BibitemShut {NoStop}%
\bibitem [{\citenamefont {Orvananos}\ \emph {et~al.}(2014)\citenamefont
  {Orvananos}, \citenamefont {Ferguson}, \citenamefont {Yu}, \citenamefont
  {Bazant},\ and\ \citenamefont {Thornton}}]{orvananos2014particle}%
  \BibitemOpen
  \bibfield  {author} {\bibinfo {author} {\bibfnamefont {B.}~\bibnamefont
  {Orvananos}}, \bibinfo {author} {\bibfnamefont {T.~R.}\ \bibnamefont
  {Ferguson}}, \bibinfo {author} {\bibfnamefont {H.-C.}\ \bibnamefont {Yu}},
  \bibinfo {author} {\bibfnamefont {M.~Z.}\ \bibnamefont {Bazant}}, \ and\
  \bibinfo {author} {\bibfnamefont {K.}~\bibnamefont {Thornton}},\ }\bibfield
  {title} {\enquote {\bibinfo {title} {Particle-level modeling of the
  charge-discharge behavior of nanoparticulate phase-separating li-ion battery
  electrodes},}\ }\href@noop {} {\bibfield  {journal} {\bibinfo  {journal} {J.
  Electrochem. Soc.}\ }\textbf {\bibinfo {volume} {161}},\ \bibinfo {pages}
  {A535} (\bibinfo {year} {2014})}\BibitemShut {NoStop}%
\bibitem [{\citenamefont {Li}\ \emph {et~al.}(2014)\citenamefont {Li},
  \citenamefont {El~Gabaly}, \citenamefont {Ferguson}, \citenamefont {Smith},
  \citenamefont {Bartelt}, \citenamefont {Sugar}, \citenamefont {Fenton},
  \citenamefont {Cogswell}, \citenamefont {Kilcoyne}, \citenamefont
  {Tyliszczak}, \citenamefont {Bazant},\ and\ \citenamefont
  {Chueh}}]{li2014current}%
  \BibitemOpen
  \bibfield  {author} {\bibinfo {author} {\bibfnamefont {Y.}~\bibnamefont
  {Li}}, \bibinfo {author} {\bibfnamefont {F.}~\bibnamefont {El~Gabaly}},
  \bibinfo {author} {\bibfnamefont {T.~R.}\ \bibnamefont {Ferguson}}, \bibinfo
  {author} {\bibfnamefont {R.~B.}\ \bibnamefont {Smith}}, \bibinfo {author}
  {\bibfnamefont {N.~C.}\ \bibnamefont {Bartelt}}, \bibinfo {author}
  {\bibfnamefont {J.~D.}\ \bibnamefont {Sugar}}, \bibinfo {author}
  {\bibfnamefont {K.~R.}\ \bibnamefont {Fenton}}, \bibinfo {author}
  {\bibfnamefont {D.~A.}\ \bibnamefont {Cogswell}}, \bibinfo {author}
  {\bibfnamefont {A.~D.}\ \bibnamefont {Kilcoyne}}, \bibinfo {author}
  {\bibfnamefont {T.}~\bibnamefont {Tyliszczak}}, \bibinfo {author}
  {\bibfnamefont {M.~Z.}\ \bibnamefont {Bazant}}, \ and\ \bibinfo {author}
  {\bibfnamefont {W.~C.}\ \bibnamefont {Chueh}},\ }\bibfield  {title} {\enquote
  {\bibinfo {title} {Current-induced transition from particle-by-particle to
  concurrent intercalation in phase-separating battery electrodes},}\
  }\href@noop {} {\bibfield  {journal} {\bibinfo  {journal} {Nature materials}\
  }\textbf {\bibinfo {volume} {13}},\ \bibinfo {pages} {1149--1156} (\bibinfo
  {year} {2014})}\BibitemShut {NoStop}%
\bibitem [{\citenamefont {Christensen}\ and\ \citenamefont
  {Newman}(2006)}]{christensen2006stress}%
  \BibitemOpen
  \bibfield  {author} {\bibinfo {author} {\bibfnamefont {J.}~\bibnamefont
  {Christensen}}\ and\ \bibinfo {author} {\bibfnamefont {J.}~\bibnamefont
  {Newman}},\ }\bibfield  {title} {\enquote {\bibinfo {title} {Stress
  generation and fracture in lithium insertion materials},}\ }\href@noop {}
  {\bibfield  {journal} {\bibinfo  {journal} {Journal of Solid State
  Electrochemistry}\ }\textbf {\bibinfo {volume} {10}},\ \bibinfo {pages}
  {293--319} (\bibinfo {year} {2006})}\BibitemShut {NoStop}%
\bibitem [{\citenamefont {Woodford}, \citenamefont {Chiang},\ and\
  \citenamefont {Carter}(2010)}]{woodford2010electrochemical}%
  \BibitemOpen
  \bibfield  {author} {\bibinfo {author} {\bibfnamefont {W.~H.}\ \bibnamefont
  {Woodford}}, \bibinfo {author} {\bibfnamefont {Y.-M.}\ \bibnamefont
  {Chiang}}, \ and\ \bibinfo {author} {\bibfnamefont {W.~C.}\ \bibnamefont
  {Carter}},\ }\bibfield  {title} {\enquote {\bibinfo {title}
  {``electrochemical shock'' of intercalation electrodes: a fracture mechanics
  analysis},}\ }\href@noop {} {\bibfield  {journal} {\bibinfo  {journal} {J.
  Electrochem. Soc.}\ }\textbf {\bibinfo {volume} {157}},\ \bibinfo {pages}
  {A1052} (\bibinfo {year} {2010})}\BibitemShut {NoStop}%
\bibitem [{\citenamefont {Qu}\ \emph {et~al.}(2018)\citenamefont {Qu},
  \citenamefont {Campbell}, \citenamefont {Hemmatifar}, \citenamefont {Knipe},
  \citenamefont {Loeb}, \citenamefont {Reidy}, \citenamefont {Hubert},
  \citenamefont {Stadermann},\ and\ \citenamefont
  {Santiago}}]{qu_charging_2018}%
  \BibitemOpen
  \bibfield  {author} {\bibinfo {author} {\bibfnamefont {Y.}~\bibnamefont
  {Qu}}, \bibinfo {author} {\bibfnamefont {P.~G.}\ \bibnamefont {Campbell}},
  \bibinfo {author} {\bibfnamefont {A.}~\bibnamefont {Hemmatifar}}, \bibinfo
  {author} {\bibfnamefont {J.~M.}\ \bibnamefont {Knipe}}, \bibinfo {author}
  {\bibfnamefont {C.~K.}\ \bibnamefont {Loeb}}, \bibinfo {author}
  {\bibfnamefont {J.~J.}\ \bibnamefont {Reidy}}, \bibinfo {author}
  {\bibfnamefont {M.~A.}\ \bibnamefont {Hubert}}, \bibinfo {author}
  {\bibfnamefont {M.}~\bibnamefont {Stadermann}}, \ and\ \bibinfo {author}
  {\bibfnamefont {J.~G.}\ \bibnamefont {Santiago}},\ }\bibfield  {title}
  {\enquote {\bibinfo {title} {Charging and {Transport} {Dynamics} of a
  {Flow}-{Through} {Electrode} {Capacitive} {Deionization} {System}},}\ }\href
  {\doibase 10.1021/acs.jpcb.7b09168} {\bibfield  {journal} {\bibinfo
  {journal} {J. Phys. Chem. B}\ }\textbf {\bibinfo {volume} {122}},\ \bibinfo
  {pages} {240--249} (\bibinfo {year} {2018})},\ \bibinfo {note} {publisher:
  American Chemical Society}\BibitemShut {NoStop}%
\bibitem [{\citenamefont {Baboukani}\ \emph {et~al.}(2019)\citenamefont
  {Baboukani}, \citenamefont {Khakpour}, \citenamefont {Drozd}, \citenamefont
  {Allagui},\ and\ \citenamefont {Wang}}]{JMCA}%
  \BibitemOpen
  \bibfield  {author} {\bibinfo {author} {\bibfnamefont {A.~R.}\ \bibnamefont
  {Baboukani}}, \bibinfo {author} {\bibfnamefont {I.}~\bibnamefont {Khakpour}},
  \bibinfo {author} {\bibfnamefont {V.}~\bibnamefont {Drozd}}, \bibinfo
  {author} {\bibfnamefont {A.}~\bibnamefont {Allagui}}, \ and\ \bibinfo
  {author} {\bibfnamefont {C.}~\bibnamefont {Wang}},\ }\bibfield  {title}
  {\enquote {\bibinfo {title} {Single-step exfoliation of black phosphorus and
  deposition of phosphorene via bipolar electrochemistry for capacitive energy
  storage application},}\ }\href@noop {} {\bibfield  {journal} {\bibinfo
  {journal} {J. Mater. Chem. A}\ }\textbf {\bibinfo {volume} {7}} (\bibinfo
  {year} {2019})}\BibitemShut {NoStop}%
\bibitem [{\citenamefont {Allagui}\ \emph {et~al.}(2017)\citenamefont
  {Allagui}, \citenamefont {Said}, \citenamefont {Abdelkareem}, \citenamefont
  {Elwakil}, \citenamefont {Yang},\ and\ \citenamefont {Alawadhi}}]{2017-3}%
  \BibitemOpen
  \bibfield  {author} {\bibinfo {author} {\bibfnamefont {A.}~\bibnamefont
  {Allagui}}, \bibinfo {author} {\bibfnamefont {Z.}~\bibnamefont {Said}},
  \bibinfo {author} {\bibfnamefont {M.~A.}\ \bibnamefont {Abdelkareem}},
  \bibinfo {author} {\bibfnamefont {A.~S.}\ \bibnamefont {Elwakil}}, \bibinfo
  {author} {\bibfnamefont {M.}~\bibnamefont {Yang}}, \ and\ \bibinfo {author}
  {\bibfnamefont {H.}~\bibnamefont {Alawadhi}},\ }\bibfield  {title} {\enquote
  {\bibinfo {title} {{DC} and {AC} performance of graphite films
  supercapacitors prepared by contact glow discharge electrolysis},}\ }\href
  {\doibase 10.1149/2.1161712jes} {\bibfield  {journal} {\bibinfo  {journal}
  {J. Electrochem. Soc.}\ }\textbf {\bibinfo {volume} {164}},\ \bibinfo {pages}
  {A2539--A2546} (\bibinfo {year} {2017})}\BibitemShut {NoStop}%
\bibitem [{\citenamefont {Khakpour}\ \emph {et~al.}(2019)\citenamefont
  {Khakpour}, \citenamefont {Baboukani}, \citenamefont {Allagui},\ and\
  \citenamefont {Wang}}]{ACSApplEnergyMater}%
  \BibitemOpen
  \bibfield  {author} {\bibinfo {author} {\bibfnamefont {I.}~\bibnamefont
  {Khakpour}}, \bibinfo {author} {\bibfnamefont {A.~R.}\ \bibnamefont
  {Baboukani}}, \bibinfo {author} {\bibfnamefont {A.}~\bibnamefont {Allagui}},
  \ and\ \bibinfo {author} {\bibfnamefont {C.}~\bibnamefont {Wang}},\
  }\bibfield  {title} {\enquote {\bibinfo {title} {Bipolar exfoliation and
  in-situ deposition of high-quality graphene for supercapacitor
  application},}\ }\href@noop {} {\bibfield  {journal} {\bibinfo  {journal}
  {ACS Appl. Energy Mater.}\ }\textbf {\bibinfo {volume} {2}},\ \bibinfo
  {pages} {4813--4820} (\bibinfo {year} {2019})}\BibitemShut {NoStop}%
\bibitem [{\citenamefont {Zhang}\ \emph {et~al.}(2019)\citenamefont {Zhang},
  \citenamefont {Allagui}, \citenamefont {Elwakil}, \citenamefont {Nassef},
  \citenamefont {Rezk}, \citenamefont {Chengi},\ and\ \citenamefont
  {Choy}}]{orgElectronics}%
  \BibitemOpen
  \bibfield  {author} {\bibinfo {author} {\bibfnamefont {D.}~\bibnamefont
  {Zhang}}, \bibinfo {author} {\bibfnamefont {A.}~\bibnamefont {Allagui}},
  \bibinfo {author} {\bibfnamefont {A.~S.}\ \bibnamefont {Elwakil}}, \bibinfo
  {author} {\bibfnamefont {A.~M.}\ \bibnamefont {Nassef}}, \bibinfo {author}
  {\bibfnamefont {H.}~\bibnamefont {Rezk}}, \bibinfo {author} {\bibfnamefont
  {J.}~\bibnamefont {Chengi}}, \ and\ \bibinfo {author} {\bibfnamefont {W.~C.}\
  \bibnamefont {Choy}},\ }\bibfield  {title} {\enquote {\bibinfo {title} {On
  the modeling of dispersive transient photocurrent response of organic solar
  cells},}\ }\href@noop {} {\bibfield  {journal} {\bibinfo  {journal} {Org.
  Electron.}\ }\textbf {\bibinfo {volume} {70}},\ \bibinfo {pages} {42--47}
  (\bibinfo {year} {2019})}\BibitemShut {NoStop}%
\bibitem [{\citenamefont {Zhang}\ \emph {et~al.}(2020)\citenamefont {Zhang},
  \citenamefont {Allagui}, \citenamefont {Elwakil}, \citenamefont {Yan},\ and\
  \citenamefont {Lu}}]{oe2}%
  \BibitemOpen
  \bibfield  {author} {\bibinfo {author} {\bibfnamefont {D.}~\bibnamefont
  {Zhang}}, \bibinfo {author} {\bibfnamefont {A.}~\bibnamefont {Allagui}},
  \bibinfo {author} {\bibfnamefont {A.~S.}\ \bibnamefont {Elwakil}}, \bibinfo
  {author} {\bibfnamefont {Z.}~\bibnamefont {Yan}}, \ and\ \bibinfo {author}
  {\bibfnamefont {H.}~\bibnamefont {Lu}},\ }\bibfield  {title} {\enquote
  {\bibinfo {title} {Active circuit model of low-frequency behavior in
  perovskite solar cells},}\ }\href@noop {} {\bibfield  {journal} {\bibinfo
  {journal} {Org. Electron.}\ }\textbf {\bibinfo {volume} {85}},\ \bibinfo
  {pages} {105804} (\bibinfo {year} {2020})}\BibitemShut {NoStop}%
\bibitem [{\citenamefont {Hilfer}(1996)}]{hilfer1996transport}%
  \BibitemOpen
  \bibfield  {author} {\bibinfo {author} {\bibfnamefont {R.}~\bibnamefont
  {Hilfer}},\ }\bibfield  {title} {\enquote {\bibinfo {title} {Transport and
  relaxation phenomena in porous media},}\ }\href@noop {} {\bibfield  {journal}
  {\bibinfo  {journal} {Advances in chemical physics}\ }\textbf {\bibinfo
  {volume} {92}},\ \bibinfo {pages} {299--424} (\bibinfo {year}
  {1996})}\BibitemShut {NoStop}%
\bibitem [{\citenamefont {Prehal}\ \emph {et~al.}(2015)\citenamefont {Prehal},
  \citenamefont {Weingarth}, \citenamefont {Perre}, \citenamefont {Lechner},
  \citenamefont {Amenitsch}, \citenamefont {Paris},\ and\ \citenamefont
  {Presser}}]{C5EE00488H}%
  \BibitemOpen
  \bibfield  {author} {\bibinfo {author} {\bibfnamefont {C.}~\bibnamefont
  {Prehal}}, \bibinfo {author} {\bibfnamefont {D.}~\bibnamefont {Weingarth}},
  \bibinfo {author} {\bibfnamefont {E.}~\bibnamefont {Perre}}, \bibinfo
  {author} {\bibfnamefont {R.~T.}\ \bibnamefont {Lechner}}, \bibinfo {author}
  {\bibfnamefont {H.}~\bibnamefont {Amenitsch}}, \bibinfo {author}
  {\bibfnamefont {O.}~\bibnamefont {Paris}}, \ and\ \bibinfo {author}
  {\bibfnamefont {V.}~\bibnamefont {Presser}},\ }\bibfield  {title} {\enquote
  {\bibinfo {title} {Tracking the structural arrangement of ions in carbon
  supercapacitor nanopores using in situ small-angle x-ray scattering},}\
  }\href {\doibase 10.1039/C5EE00488H} {\bibfield  {journal} {\bibinfo
  {journal} {Energy Environ. Sci.}\ }\textbf {\bibinfo {volume} {8}},\ \bibinfo
  {pages} {1725--1735} (\bibinfo {year} {2015})}\BibitemShut {NoStop}%
\bibitem [{\citenamefont {Prehal}\ \emph {et~al.}(2017)\citenamefont {Prehal},
  \citenamefont {Koczwara}, \citenamefont {J{\"a}ckel}, \citenamefont
  {Amenitsch}, \citenamefont {Presser},\ and\ \citenamefont
  {Paris}}]{C7CP00736A}%
  \BibitemOpen
  \bibfield  {author} {\bibinfo {author} {\bibfnamefont {C.}~\bibnamefont
  {Prehal}}, \bibinfo {author} {\bibfnamefont {C.}~\bibnamefont {Koczwara}},
  \bibinfo {author} {\bibfnamefont {N.}~\bibnamefont {J{\"a}ckel}}, \bibinfo
  {author} {\bibfnamefont {H.}~\bibnamefont {Amenitsch}}, \bibinfo {author}
  {\bibfnamefont {V.}~\bibnamefont {Presser}}, \ and\ \bibinfo {author}
  {\bibfnamefont {O.}~\bibnamefont {Paris}},\ }\bibfield  {title} {\enquote
  {\bibinfo {title} {A carbon nanopore model to quantify structure and kinetics
  of ion electrosorption with in situ small-angle x-ray scattering},}\ }\href
  {\doibase 10.1039/C7CP00736A} {\bibfield  {journal} {\bibinfo  {journal}
  {Phys. Chem. Chem. Phys.}\ }\textbf {\bibinfo {volume} {19}},\ \bibinfo
  {pages} {15549--15561} (\bibinfo {year} {2017})}\BibitemShut {NoStop}%
\bibitem [{\citenamefont {Prehal}\ \emph {et~al.}(2018)\citenamefont {Prehal},
  \citenamefont {Koczwara}, \citenamefont {Amenitsch}, \citenamefont
  {Presser},\ and\ \citenamefont {Paris}}]{prehal_salt_2018}%
  \BibitemOpen
  \bibfield  {author} {\bibinfo {author} {\bibfnamefont {C.}~\bibnamefont
  {Prehal}}, \bibinfo {author} {\bibfnamefont {C.}~\bibnamefont {Koczwara}},
  \bibinfo {author} {\bibfnamefont {H.}~\bibnamefont {Amenitsch}}, \bibinfo
  {author} {\bibfnamefont {V.}~\bibnamefont {Presser}}, \ and\ \bibinfo
  {author} {\bibfnamefont {O.}~\bibnamefont {Paris}},\ }\bibfield  {title}
  {\enquote {\bibinfo {title} {Salt concentration and charging velocity
  determine ion charge storage mechanism in nanoporous supercapacitors},}\
  }\href {\doibase 10.1038/s41467-018-06612-4} {\bibfield  {journal} {\bibinfo
  {journal} {Nature Communications}\ }\textbf {\bibinfo {volume} {9}},\
  \bibinfo {pages} {4145} (\bibinfo {year} {2018})}\BibitemShut {NoStop}%
\bibitem [{\citenamefont {Haubold}\ and\ \citenamefont
  {Mathai}(2000)}]{haubold2000fractional}%
  \BibitemOpen
  \bibfield  {author} {\bibinfo {author} {\bibfnamefont {H.~J.}\ \bibnamefont
  {Haubold}}\ and\ \bibinfo {author} {\bibfnamefont {A.~M.}\ \bibnamefont
  {Mathai}},\ }\bibfield  {title} {\enquote {\bibinfo {title} {The fractional
  kinetic equation and thermonuclear functions},}\ }\href@noop {} {\bibfield
  {journal} {\bibinfo  {journal} {Astrophysics and Space Science}\ }\textbf
  {\bibinfo {volume} {273}},\ \bibinfo {pages} {53--63} (\bibinfo {year}
  {2000})}\BibitemShut {NoStop}%
\bibitem [{\citenamefont {Mainardi}(1996)}]{mainardi1996fractional}%
  \BibitemOpen
  \bibfield  {author} {\bibinfo {author} {\bibfnamefont {F.}~\bibnamefont
  {Mainardi}},\ }\bibfield  {title} {\enquote {\bibinfo {title} {Fractional
  relaxation-oscillation and fractional diffusion-wave phenomena},}\
  }\href@noop {} {\bibfield  {journal} {\bibinfo  {journal} {Chaos, Solitons \&
  Fractals}\ }\textbf {\bibinfo {volume} {7}},\ \bibinfo {pages} {1461--1477}
  (\bibinfo {year} {1996})}\BibitemShut {NoStop}%
\bibitem [{\citenamefont {Mathai}\ and\ \citenamefont
  {Haubold}(2007)}]{mathai2007pathway}%
  \BibitemOpen
  \bibfield  {author} {\bibinfo {author} {\bibfnamefont {A.}~\bibnamefont
  {Mathai}}\ and\ \bibinfo {author} {\bibfnamefont {H.~J.}\ \bibnamefont
  {Haubold}},\ }\bibfield  {title} {\enquote {\bibinfo {title} {Pathway model,
  superstatistics, tsallis statistics, and a generalized measure of entropy},}\
  }\href@noop {} {\bibfield  {journal} {\bibinfo  {journal} {Physica A:
  Statistical Mechanics and its Applications}\ }\textbf {\bibinfo {volume}
  {375}},\ \bibinfo {pages} {110--122} (\bibinfo {year} {2007})}\BibitemShut
  {NoStop}%
\bibitem [{\citenamefont {Mathai}, \citenamefont {Saxena},\ and\ \citenamefont
  {Haubold}(2009)}]{mathai2009h}%
  \BibitemOpen
  \bibfield  {author} {\bibinfo {author} {\bibfnamefont {A.~M.}\ \bibnamefont
  {Mathai}}, \bibinfo {author} {\bibfnamefont {R.~K.}\ \bibnamefont {Saxena}},
  \ and\ \bibinfo {author} {\bibfnamefont {H.~J.}\ \bibnamefont {Haubold}},\
  }\href@noop {} {\emph {\bibinfo {title} {The H-function: theory and
  applications}}}\ (\bibinfo  {publisher} {Springer Science \& Business
  Media},\ \bibinfo {year} {2009})\BibitemShut {NoStop}%
\bibitem [{\citenamefont {Saxena}, \citenamefont {Mathai},\ and\ \citenamefont
  {Haubold}(2004)}]{saxena2004generalized}%
  \BibitemOpen
  \bibfield  {author} {\bibinfo {author} {\bibfnamefont {R.}~\bibnamefont
  {Saxena}}, \bibinfo {author} {\bibfnamefont {A.}~\bibnamefont {Mathai}}, \
  and\ \bibinfo {author} {\bibfnamefont {H.}~\bibnamefont {Haubold}},\
  }\bibfield  {title} {\enquote {\bibinfo {title} {On generalized fractional
  kinetic equations},}\ }\href@noop {} {\bibfield  {journal} {\bibinfo
  {journal} {Physica A: Statistical Mechanics and its Applications}\ }\textbf
  {\bibinfo {volume} {344}},\ \bibinfo {pages} {657--664} (\bibinfo {year}
  {2004})}\BibitemShut {NoStop}%
\bibitem [{\citenamefont {Beck}(2004)}]{beck2004superstatistics}%
  \BibitemOpen
  \bibfield  {author} {\bibinfo {author} {\bibfnamefont {C.}~\bibnamefont
  {Beck}},\ }\bibfield  {title} {\enquote {\bibinfo {title} {Superstatistics:
  theory and applications},}\ }\href@noop {} {\bibfield  {journal} {\bibinfo
  {journal} {Continuum Mech. Thermodyn.}\ }\textbf {\bibinfo {volume} {16}},\
  \bibinfo {pages} {293--304} (\bibinfo {year} {2004})}\BibitemShut {NoStop}%
\bibitem [{\citenamefont {Beck}\ and\ \citenamefont
  {Cohen}(2003)}]{beck2003superstatistics}%
  \BibitemOpen
  \bibfield  {author} {\bibinfo {author} {\bibfnamefont {C.}~\bibnamefont
  {Beck}}\ and\ \bibinfo {author} {\bibfnamefont {E.~G.}\ \bibnamefont
  {Cohen}},\ }\bibfield  {title} {\enquote {\bibinfo {title}
  {Superstatistics},}\ }\href@noop {} {\bibfield  {journal} {\bibinfo
  {journal} {Physica A}\ }\textbf {\bibinfo {volume} {322}},\ \bibinfo {pages}
  {267--275} (\bibinfo {year} {2003})}\BibitemShut {NoStop}%
\bibitem [{\citenamefont {Drazer}\ and\ \citenamefont
  {Zanette}(1999)}]{drazer1999experimental}%
  \BibitemOpen
  \bibfield  {author} {\bibinfo {author} {\bibfnamefont {G.}~\bibnamefont
  {Drazer}}\ and\ \bibinfo {author} {\bibfnamefont {D.~H.}\ \bibnamefont
  {Zanette}},\ }\bibfield  {title} {\enquote {\bibinfo {title} {Experimental
  evidence of power-law trapping-time distributions in porous media},}\
  }\href@noop {} {\bibfield  {journal} {\bibinfo  {journal} {Physical Review
  E}\ }\textbf {\bibinfo {volume} {60}},\ \bibinfo {pages} {5858} (\bibinfo
  {year} {1999})}\BibitemShut {NoStop}%
\bibitem [{\citenamefont {Allagui}, \citenamefont {Zhang},\ and\ \citenamefont
  {Elwakil}(2018)}]{memoryAPL}%
  \BibitemOpen
  \bibfield  {author} {\bibinfo {author} {\bibfnamefont {A.}~\bibnamefont
  {Allagui}}, \bibinfo {author} {\bibfnamefont {D.}~\bibnamefont {Zhang}}, \
  and\ \bibinfo {author} {\bibfnamefont {A.~S.}\ \bibnamefont {Elwakil}},\
  }\bibfield  {title} {\enquote {\bibinfo {title} {Short-term memory in
  electric double-layer capacitors},}\ }\href@noop {} {\bibfield  {journal}
  {\bibinfo  {journal} {Appl. Phys. Lett.}\ }\textbf {\bibinfo {volume}
  {113}},\ \bibinfo {pages} {253901--5} (\bibinfo {year} {2018})}\BibitemShut
  {NoStop}%
\bibitem [{\citenamefont {Dattoli}\ \emph {et~al.}(2014)\citenamefont
  {Dattoli}, \citenamefont {G{\'o}rska}, \citenamefont {Horzela},\ and\
  \citenamefont {Penson}}]{dattoli2014photoluminescence}%
  \BibitemOpen
  \bibfield  {author} {\bibinfo {author} {\bibfnamefont {G.}~\bibnamefont
  {Dattoli}}, \bibinfo {author} {\bibfnamefont {K.}~\bibnamefont {G{\'o}rska}},
  \bibinfo {author} {\bibfnamefont {A.}~\bibnamefont {Horzela}}, \ and\
  \bibinfo {author} {\bibfnamefont {K.}~\bibnamefont {Penson}},\ }\bibfield
  {title} {\enquote {\bibinfo {title} {Photoluminescence decay of silicon
  nanocrystals and l{\'e}vy stable distributions},}\ }\href@noop {} {\bibfield
  {journal} {\bibinfo  {journal} {Physics Letters A}\ }\textbf {\bibinfo
  {volume} {378}},\ \bibinfo {pages} {2201--2205} (\bibinfo {year}
  {2014})}\BibitemShut {NoStop}%
\bibitem [{\citenamefont {Metzler}\ and\ \citenamefont
  {Klafter}(2000)}]{metzler2000random}%
  \BibitemOpen
  \bibfield  {author} {\bibinfo {author} {\bibfnamefont {R.}~\bibnamefont
  {Metzler}}\ and\ \bibinfo {author} {\bibfnamefont {J.}~\bibnamefont
  {Klafter}},\ }\bibfield  {title} {\enquote {\bibinfo {title} {The random
  walk's guide to anomalous diffusion: a fractional dynamics approach},}\
  }\href@noop {} {\bibfield  {journal} {\bibinfo  {journal} {Physics reports}\
  }\textbf {\bibinfo {volume} {339}},\ \bibinfo {pages} {1--77} (\bibinfo
  {year} {2000})}\BibitemShut {NoStop}%
\bibitem [{\citenamefont {Chechkin}, \citenamefont {Gorenflo},\ and\
  \citenamefont {Sokolov}(2002)}]{chechkin2002retarding}%
  \BibitemOpen
  \bibfield  {author} {\bibinfo {author} {\bibfnamefont {A.}~\bibnamefont
  {Chechkin}}, \bibinfo {author} {\bibfnamefont {R.}~\bibnamefont {Gorenflo}},
  \ and\ \bibinfo {author} {\bibfnamefont {I.}~\bibnamefont {Sokolov}},\
  }\bibfield  {title} {\enquote {\bibinfo {title} {Retarding subdiffusion and
  accelerating superdiffusion governed by distributed-order fractional
  diffusion equations},}\ }\href@noop {} {\bibfield  {journal} {\bibinfo
  {journal} {Physical Review E}\ }\textbf {\bibinfo {volume} {66}},\ \bibinfo
  {pages} {046129} (\bibinfo {year} {2002})}\BibitemShut {NoStop}%
\bibitem [{\citenamefont {Ribeiro}\ and\ \citenamefont
  {Potiguar}(2016)}]{ribeiro2016active}%
  \BibitemOpen
  \bibfield  {author} {\bibinfo {author} {\bibfnamefont {H.}~\bibnamefont
  {Ribeiro}}\ and\ \bibinfo {author} {\bibfnamefont {F.}~\bibnamefont
  {Potiguar}},\ }\bibfield  {title} {\enquote {\bibinfo {title} {Active matter
  in lateral parabolic confinement: From subdiffusion to superdiffusion},}\
  }\href@noop {} {\bibfield  {journal} {\bibinfo  {journal} {Physica A:
  Statistical Mechanics and its Applications}\ }\textbf {\bibinfo {volume}
  {462}},\ \bibinfo {pages} {1294--1300} (\bibinfo {year} {2016})}\BibitemShut
  {NoStop}%
\bibitem [{\citenamefont {Kneller}\ and\ \citenamefont
  {Hinsen}(2004)}]{doi:10.1063/1.1806134}%
  \BibitemOpen
  \bibfield  {author} {\bibinfo {author} {\bibfnamefont {G.~R.}\ \bibnamefont
  {Kneller}}\ and\ \bibinfo {author} {\bibfnamefont {K.}~\bibnamefont
  {Hinsen}},\ }\bibfield  {title} {\enquote {\bibinfo {title} {Fractional
  brownian dynamics in proteins},}\ }\href {\doibase 10.1063/1.1806134}
  {\bibfield  {journal} {\bibinfo  {journal} {The Journal of Chemical Physics}\
  }\textbf {\bibinfo {volume} {121}},\ \bibinfo {pages} {10278--10283}
  (\bibinfo {year} {2004})},\ \Eprint
  {http://arxiv.org/abs/https://doi.org/10.1063/1.1806134}
  {https://doi.org/10.1063/1.1806134} \BibitemShut {NoStop}%
\bibitem [{\citenamefont {Allagui}\ and\ \citenamefont
  {Elwakil}(2021)}]{allagui2021possibility}%
  \BibitemOpen
  \bibfield  {author} {\bibinfo {author} {\bibfnamefont {A.}~\bibnamefont
  {Allagui}}\ and\ \bibinfo {author} {\bibfnamefont {A.~S.}\ \bibnamefont
  {Elwakil}},\ }\bibfield  {title} {\enquote {\bibinfo {title} {Possibility of
  information encoding/decoding using the memory effect in fractional-order
  capacitive devices},}\ }\href@noop {} {\bibfield  {journal} {\bibinfo
  {journal} {Sci. Rep.}\ }\textbf {\bibinfo {volume} {11}},\ \bibinfo {pages}
  {1--7} (\bibinfo {year} {2021})}\BibitemShut {NoStop}%
\bibitem [{\citenamefont {Allagui}\ \emph {et~al.}(2020)\citenamefont
  {Allagui}, \citenamefont {Zhang}, \citenamefont {Khakpour}, \citenamefont
  {Elwakil},\ and\ \citenamefont {Wang}}]{memQ}%
  \BibitemOpen
  \bibfield  {author} {\bibinfo {author} {\bibfnamefont {A.}~\bibnamefont
  {Allagui}}, \bibinfo {author} {\bibfnamefont {D.}~\bibnamefont {Zhang}},
  \bibinfo {author} {\bibfnamefont {I.}~\bibnamefont {Khakpour}}, \bibinfo
  {author} {\bibfnamefont {A.~S.}\ \bibnamefont {Elwakil}}, \ and\ \bibinfo
  {author} {\bibfnamefont {C.}~\bibnamefont {Wang}},\ }\bibfield  {title}
  {\enquote {\bibinfo {title} {Quantification of memory in fractional-order
  capacitors},}\ }\href@noop {} {\bibfield  {journal} {\bibinfo  {journal} {J.
  Phys. D}\ }\textbf {\bibinfo {volume} {53}} (\bibinfo {year}
  {2020})}\BibitemShut {NoStop}%
\bibitem [{\citenamefont {Koszto{\l}owicz}\ and\ \citenamefont
  {Dutkiewicz}(2021)}]{kosztolowicz2021subdiffusion}%
  \BibitemOpen
  \bibfield  {author} {\bibinfo {author} {\bibfnamefont {T.}~\bibnamefont
  {Koszto{\l}owicz}}\ and\ \bibinfo {author} {\bibfnamefont {A.}~\bibnamefont
  {Dutkiewicz}},\ }\bibfield  {title} {\enquote {\bibinfo {title} {Subdiffusion
  equation with caputo fractional derivative with respect to another
  function},}\ }\href@noop {} {\bibfield  {journal} {\bibinfo  {journal} {arXiv
  preprint arXiv:2104.14918}\ } (\bibinfo {year} {2021})}\BibitemShut {NoStop}%
\bibitem [{\citenamefont {Huang}\ and\ \citenamefont
  {Liu}(2005)}]{huang2005space}%
  \BibitemOpen
  \bibfield  {author} {\bibinfo {author} {\bibfnamefont {F.}~\bibnamefont
  {Huang}}\ and\ \bibinfo {author} {\bibfnamefont {F.}~\bibnamefont {Liu}},\
  }\bibfield  {title} {\enquote {\bibinfo {title} {The space-time fractional
  diffusion equation with caputo derivatives},}\ }\href@noop {} {\bibfield
  {journal} {\bibinfo  {journal} {Journal of Applied Mathematics and
  Computing}\ }\textbf {\bibinfo {volume} {19}},\ \bibinfo {pages} {179--190}
  (\bibinfo {year} {2005})}\BibitemShut {NoStop}%
\bibitem [{\citenamefont {Prabhakar}(1971)}]{prabhakar1971singular}%
  \BibitemOpen
  \bibfield  {author} {\bibinfo {author} {\bibfnamefont {T.~R.}\ \bibnamefont
  {Prabhakar}},\ }\bibfield  {title} {\enquote {\bibinfo {title} {A singular
  integral equation with a generalized mittag leffler function in the
  kernel},}\ }\href@noop {} {\bibfield  {journal} {\bibinfo  {journal}
  {Yokohama Mathematical Journal}\ }\textbf {\bibinfo {volume} {19}},\ \bibinfo
  {pages} {7--15} (\bibinfo {year} {1971})}\BibitemShut {NoStop}%
\bibitem [{\citenamefont {Mainardi}(2014)}]{mainardi2014some}%
  \BibitemOpen
  \bibfield  {author} {\bibinfo {author} {\bibfnamefont {F.}~\bibnamefont
  {Mainardi}},\ }\bibfield  {title} {\enquote {\bibinfo {title} {On some
  properties of the mittag-leffler function $e_{\alpha}(-t^{\alpha})$,
  completely monotone for $t> 0$ with $0<\alpha< 1$},}\ }\href@noop {}
  {\bibfield  {journal} {\bibinfo  {journal} {Discrete \& Continuous Dynamical
  Systems-B}\ }\textbf {\bibinfo {volume} {19}},\ \bibinfo {pages} {2267}
  (\bibinfo {year} {2014})}\BibitemShut {NoStop}%
\bibitem [{\citenamefont {Lemes}, \citenamefont {dos Santos},\ and\
  \citenamefont {Braga}(2016)}]{lemes2016generalized}%
  \BibitemOpen
  \bibfield  {author} {\bibinfo {author} {\bibfnamefont {N.~H.}\ \bibnamefont
  {Lemes}}, \bibinfo {author} {\bibfnamefont {J.~P.~C.}\ \bibnamefont {dos
  Santos}}, \ and\ \bibinfo {author} {\bibfnamefont {J.~P.}\ \bibnamefont
  {Braga}},\ }\bibfield  {title} {\enquote {\bibinfo {title} {A generalized
  mittag-leffler function to describe nonexponential chemical effects},}\
  }\href@noop {} {\bibfield  {journal} {\bibinfo  {journal} {Applied
  Mathematical Modelling}\ }\textbf {\bibinfo {volume} {40}},\ \bibinfo {pages}
  {7971--7976} (\bibinfo {year} {2016})}\BibitemShut {NoStop}%
\bibitem [{\citenamefont {De~Oliveira}, \citenamefont {Mainardi},\ and\
  \citenamefont {Vaz}(2011)}]{de2011models}%
  \BibitemOpen
  \bibfield  {author} {\bibinfo {author} {\bibfnamefont {E.~C.}\ \bibnamefont
  {De~Oliveira}}, \bibinfo {author} {\bibfnamefont {F.}~\bibnamefont
  {Mainardi}}, \ and\ \bibinfo {author} {\bibfnamefont {J.}~\bibnamefont
  {Vaz}},\ }\bibfield  {title} {\enquote {\bibinfo {title} {Models based on
  mittag-leffler functions for anomalous relaxation in dielectrics},}\
  }\href@noop {} {\bibfield  {journal} {\bibinfo  {journal} {The European
  Physical Journal Special Topics}\ }\textbf {\bibinfo {volume} {193}},\
  \bibinfo {pages} {161--171} (\bibinfo {year} {2011})}\BibitemShut {NoStop}%
\bibitem [{\citenamefont {Mathai}\ and\ \citenamefont
  {Haubold}(2017)}]{mathai2017probability}%
  \BibitemOpen
  \bibfield  {author} {\bibinfo {author} {\bibfnamefont {A.~M.}\ \bibnamefont
  {Mathai}}\ and\ \bibinfo {author} {\bibfnamefont {H.~J.}\ \bibnamefont
  {Haubold}},\ }\href@noop {} {\emph {\bibinfo {title} {Probability and
  statistics}}}\ (\bibinfo  {publisher} {De Gruyter},\ \bibinfo {year}
  {2017})\BibitemShut {NoStop}%
\bibitem [{\citenamefont {Lienhard}\ and\ \citenamefont
  {Meyer}(1967)}]{lienhard1967physical}%
  \BibitemOpen
  \bibfield  {author} {\bibinfo {author} {\bibfnamefont {J.~H.}\ \bibnamefont
  {Lienhard}}\ and\ \bibinfo {author} {\bibfnamefont {P.~L.}\ \bibnamefont
  {Meyer}},\ }\bibfield  {title} {\enquote {\bibinfo {title} {A physical basis
  for the generalized gamma distribution},}\ }\href@noop {} {\bibfield
  {journal} {\bibinfo  {journal} {Quarterly of Applied Mathematics}\ }\textbf
  {\bibinfo {volume} {25}},\ \bibinfo {pages} {330--334} (\bibinfo {year}
  {1967})}\BibitemShut {NoStop}%
\bibitem [{\citenamefont {Beck}(2006)}]{beck2006stretched}%
  \BibitemOpen
  \bibfield  {author} {\bibinfo {author} {\bibfnamefont {C.}~\bibnamefont
  {Beck}},\ }\bibfield  {title} {\enquote {\bibinfo {title} {Stretched
  exponentials from superstatistics},}\ }\href@noop {} {\bibfield  {journal}
  {\bibinfo  {journal} {Physica A: Statistical Mechanics and its Applications}\
  }\textbf {\bibinfo {volume} {365}},\ \bibinfo {pages} {96--101} (\bibinfo
  {year} {2006})}\BibitemShut {NoStop}%
\bibitem [{\citenamefont {Mathai}\ and\ \citenamefont
  {Moschopoulos}(2012)}]{mathai2012pathway}%
  \BibitemOpen
  \bibfield  {author} {\bibinfo {author} {\bibfnamefont {A.}~\bibnamefont
  {Mathai}}\ and\ \bibinfo {author} {\bibfnamefont {P.}~\bibnamefont
  {Moschopoulos}},\ }\bibfield  {title} {\enquote {\bibinfo {title} {A pathway
  idea for model building},}\ }\href@noop {} {\bibfield  {journal} {\bibinfo
  {journal} {Journal of statistics applications \& probability}\ }\textbf
  {\bibinfo {volume} {1}},\ \bibinfo {pages} {15} (\bibinfo {year}
  {2012})}\BibitemShut {NoStop}%
\bibitem [{\citenamefont {Allagui}, \citenamefont {Benaoum},\ and\
  \citenamefont {Olendski}(2021)}]{allagui2021gouy}%
  \BibitemOpen
  \bibfield  {author} {\bibinfo {author} {\bibfnamefont {A.}~\bibnamefont
  {Allagui}}, \bibinfo {author} {\bibfnamefont {H.}~\bibnamefont {Benaoum}}, \
  and\ \bibinfo {author} {\bibfnamefont {O.}~\bibnamefont {Olendski}},\
  }\bibfield  {title} {\enquote {\bibinfo {title} {On the gouy-chapman-stern
  model of the electrical double-layer structure with a generalized boltzmann
  factor},}\ }\href@noop {} {\bibfield  {journal} {\bibinfo  {journal} {Physica
  A}\ ,\ \bibinfo {pages} {126252}} (\bibinfo {year} {2021})}\BibitemShut
  {NoStop}%
\bibitem [{\citenamefont {Stanislavsky}(2007)}]{stanislavsky2007stochastic}%
  \BibitemOpen
  \bibfield  {author} {\bibinfo {author} {\bibfnamefont {A.~A.}\ \bibnamefont
  {Stanislavsky}},\ }\bibfield  {title} {\enquote {\bibinfo {title} {The
  stochastic nature of complexity evolution in the fractional systems},}\
  }\href@noop {} {\bibfield  {journal} {\bibinfo  {journal} {Chaos, Solitons \&
  Fractals}\ }\textbf {\bibinfo {volume} {34}},\ \bibinfo {pages} {51--61}
  (\bibinfo {year} {2007})}\BibitemShut {NoStop}%
\bibitem [{\citenamefont {Stanislavsky}, \citenamefont {Weron},\ and\
  \citenamefont {Trzmiel}(2010)}]{stanislavsky2010subordination}%
  \BibitemOpen
  \bibfield  {author} {\bibinfo {author} {\bibfnamefont {A.}~\bibnamefont
  {Stanislavsky}}, \bibinfo {author} {\bibfnamefont {K.}~\bibnamefont {Weron}},
  \ and\ \bibinfo {author} {\bibfnamefont {J.}~\bibnamefont {Trzmiel}},\
  }\bibfield  {title} {\enquote {\bibinfo {title} {Subordination model of
  anomalous diffusion leading to the two-power-law relaxation responses},}\
  }\href@noop {} {\bibfield  {journal} {\bibinfo  {journal} {EPL (Europhysics
  Letters)}\ }\textbf {\bibinfo {volume} {91}},\ \bibinfo {pages} {40003}
  (\bibinfo {year} {2010})}\BibitemShut {NoStop}%
\bibitem [{\citenamefont {Chechkin}\ \emph {et~al.}(2017)\citenamefont
  {Chechkin}, \citenamefont {Seno}, \citenamefont {Metzler},\ and\
  \citenamefont {Sokolov}}]{chechkin2017brownian}%
  \BibitemOpen
  \bibfield  {author} {\bibinfo {author} {\bibfnamefont {A.~V.}\ \bibnamefont
  {Chechkin}}, \bibinfo {author} {\bibfnamefont {F.}~\bibnamefont {Seno}},
  \bibinfo {author} {\bibfnamefont {R.}~\bibnamefont {Metzler}}, \ and\
  \bibinfo {author} {\bibfnamefont {I.~M.}\ \bibnamefont {Sokolov}},\
  }\bibfield  {title} {\enquote {\bibinfo {title} {Brownian yet non-gaussian
  diffusion: from superstatistics to subordination of diffusing
  diffusivities},}\ }\href@noop {} {\bibfield  {journal} {\bibinfo  {journal}
  {Physical Review X}\ }\textbf {\bibinfo {volume} {7}},\ \bibinfo {pages}
  {021002} (\bibinfo {year} {2017})}\BibitemShut {NoStop}%
\bibitem [{\citenamefont {Allagui}\ \emph {et~al.}(2018)\citenamefont
  {Allagui}, \citenamefont {Freeborn}, \citenamefont {Elwakil}, \citenamefont
  {Fouda}, \citenamefont {Maundy}, \citenamefont {Radwanh}, \citenamefont
  {Said},\ and\ \citenamefont {Abdelkareem}}]{fracorderreview}%
  \BibitemOpen
  \bibfield  {author} {\bibinfo {author} {\bibfnamefont {A.}~\bibnamefont
  {Allagui}}, \bibinfo {author} {\bibfnamefont {T.~J.}\ \bibnamefont
  {Freeborn}}, \bibinfo {author} {\bibfnamefont {A.~S.}\ \bibnamefont
  {Elwakil}}, \bibinfo {author} {\bibfnamefont {M.~E.}\ \bibnamefont {Fouda}},
  \bibinfo {author} {\bibfnamefont {B.~J.}\ \bibnamefont {Maundy}}, \bibinfo
  {author} {\bibfnamefont {A.~G.}\ \bibnamefont {Radwanh}}, \bibinfo {author}
  {\bibfnamefont {Z.}~\bibnamefont {Said}}, \ and\ \bibinfo {author}
  {\bibfnamefont {M.~A.}\ \bibnamefont {Abdelkareem}},\ }\bibfield  {title}
  {\enquote {\bibinfo {title} {Review of fractional-order electrical
  characterization of supercapacitors},}\ }\href@noop {} {\bibfield  {journal}
  {\bibinfo  {journal} {J. Power Sources}\ }\textbf {\bibinfo {volume} {400}}
  (\bibinfo {year} {2018})}\BibitemShut {NoStop}%
\bibitem [{\citenamefont {Allagui}\ \emph {et~al.}(2016)\citenamefont
  {Allagui}, \citenamefont {Elwakil}, \citenamefont {Maundy},\ and\
  \citenamefont {Freeborn}}]{eis}%
  \BibitemOpen
  \bibfield  {author} {\bibinfo {author} {\bibfnamefont {A.}~\bibnamefont
  {Allagui}}, \bibinfo {author} {\bibfnamefont {A.~S.}\ \bibnamefont
  {Elwakil}}, \bibinfo {author} {\bibfnamefont {B.~J.}\ \bibnamefont {Maundy}},
  \ and\ \bibinfo {author} {\bibfnamefont {T.~J.}\ \bibnamefont {Freeborn}},\
  }\bibfield  {title} {\enquote {\bibinfo {title} {Spectral capacitance of
  series and parallel combinations of supercapacitors},}\ }\href@noop {}
  {\bibfield  {journal} {\bibinfo  {journal} {ChemElectroChem}\ }\textbf
  {\bibinfo {volume} {3}},\ \bibinfo {pages} {1429--1436} (\bibinfo {year}
  {2016})}\BibitemShut {NoStop}%
\end{thebibliography}

%

 \end{document}